\documentclass[aps,prb,twoside,twocolumn,10pt,floatfix,showpacs,citeautoscript,superscriptaddress]{revtex4-2}

\usepackage{comment}
\usepackage{glossaries}
\usepackage{graphicx}
\usepackage[version=4]{mhchem}
\usepackage[per-mode=symbol]{siunitx}
\usepackage[dvipsnames,svgnames]{xcolor}
\usepackage{hyperref}
\hypersetup{
pdfnewwindow=true,          
colorlinks=true,            
linkcolor=DarkBlue,         
citecolor=DarkBlue,         
filecolor=DarkBlue,         
urlcolor=DarkBlue           
}

\graphicspath{{figures}}

\renewcommand{\vec}[1]{\ensuremath\mathbf{#1}}
\DeclareSIUnit\angstrom{\text{Å}}
\DeclareSIUnit\atom{\text{atom}}
\DeclareSIUnit\step{\text{step}}

\makeatletter
\let\oldtheequation\theequation
\renewcommand\tagform@[1]{\maketag@@@{\ignorespaces#1\unskip\@@italiccorr}}
\renewcommand\theequation{(\oldtheequation)}
\makeatother

\newacronym{acf}{ACF}{autocorrelation function}
\newacronym{bec}{BEC}{Born effective charge}
\newacronym{dft}{DFT}{density-functional theory}
\newacronym{fft}{FFT}{fast Fourier transform}
\newacronym{les}{LES}{latent Ewald summation}
\newacronym{llzo}{LLZO}{lithium lanthanum zirconate \ce{Li7La3Zr2O12}}
\newacronym{md}{MD}{molecular dynamics}
\newacronym{mlip}{MLIP}{machine-learned interatomic potential}
\newacronym{nac}{NAC}{non-analytic correction}
\newacronym{nep}{NEP}{neuroevolution potential}
\newacronym{nn}{NN}{neural network}
\newacronym{pimd}{PIMD}{path-integral molecular dynamics}
\newacronym{pppm}{PPPM}{particle-particle particle-mesh}
\newacronym{rmse}{RMSE}{root-mean-square error}
\newacronym{scan}{SCAN}{strongly constrained and appropriately normed semilocal density functional}
\newacronym{snes}{SNES}{separable natural evolution strategy}
\newacronym{zbl}{ZBL}{Ziegler-Biersack-Littmark}

\newcommand{\addressBohai}{College of Physical Science and Technology, Bohai University, Jinzhou, P. R. China}
\newcommand{\addressChalmersPhysics}{Chalmers University of Technology, Department of Physics, 41296 Gothenburg, Sweden}
\newcommand{\addressChalmersWISE}{Wallenberg Initiative Materials Science for Sustainability, Chalmers University of Technology, 41926 Gothenburg, Sweden}
\newcommand{\addressWestlake}{Department of Materials Science and Engineering, Westlake University, Hangzhou, Zhejiang 310030, China}
\newcommand{\addressShenzhen}{Shenzhen Key Laboratory of Micro/Nano-Porous Functional Materials (SKLPM), Department of Materials Science and Engineering, Southern University of Science and Technology, Shenzhen 518055, China}

\begin{document}

\title{qNEP: A highly efficient neuroevolution potential \texorpdfstring{\\}{ }with dynamic charges for large-scale atomistic simulations}

\author{Zheyong Fan}
\email{brucenju@gmail.com}
\thanks{These authors contributed equally to this work.}
\affiliation{\addressBohai}
\affiliation{Suzhou Laboratory, Suzhou, Jiangsu 215123, P. R. China}

\author{Benrui Tang}
\thanks{These authors contributed equally to this work.}
\affiliation{\addressBohai}

\author{Esmée Berger}
\thanks{These authors contributed equally to this work.}
\author{Ethan Berger}
\author{Erik Fransson}
\affiliation{\addressChalmersPhysics}

\author{Ke Xu}
\affiliation{\addressBohai}

\author{Zihan Yan}
\affiliation{\addressWestlake}

\author{Zhoulin Liu}
\affiliation{School of Science, Harbin Institute of Technology, Shenzhen 518055, Guangdong, P. R. China}

\author{Zichen Song}
\affiliation{\addressShenzhen}
\affiliation{Department of Materials Science and Engineering, City University of Hong Kong, Hong Kong SAR, China}

\author{Haikuan Dong}
\affiliation{\addressBohai}

\author{Shunda Chen}
\affiliation{Department of Civil and Environmental Engineering, George Washington University, Washington, DC 20052, USA}

\author{Lei Li}
\affiliation{\addressShenzhen}

\author{Ziliang Wang}
\affiliation{National Engineering Laboratory for Reducing Emissions from Coal Combustion, Shandong Key Laboratory of Green Thermal Power and Carbon Reduction, Shandong University, Jinan, Shandong, P. R. China}

\author{Yizhou Zhu}
\affiliation{\addressWestlake}

\author{Julia Wiktor}
\affiliation{\addressChalmersPhysics}

\author{Paul Erhart}
\email{erhart@chalmers.se}
\affiliation{\addressChalmersPhysics}
\affiliation{\addressChalmersWISE}

\date{\today}

\begin{abstract}
Although electrostatics can be incorporated into machine-learned interatomic potentials, existing approaches are computationally very demanding, limiting large-scale, long-time simulations of electrostatics-driven phenomena such as dielectric response, infrared activity, and field–matter coupling.
Here, we extend the neuroevolution potential (NEP), a highly efficient machine-learned interatomic potential, to a charge-aware framework (qNEP) by introducing explicit, environment-dependent partial charges.
Each ionic partial charge is represented by a neural network as a function of the local descriptor vector, analogous to the NEP site-energy model.
This formulation enables the direct prediction of the Born effective charge tensor for each ion and, consequently, the polarization.
As a result, dielectric properties, infrared spectra, and coupling to external electric fields can be evaluated within a unified framework.
We derive consistent expressions for the forces and virials that explicitly account for the position dependence of the partial charges.
The qNEP method has been implemented in the free-and-open-source GPUMD package, with support for both Ewald summation and particle–particle particle–mesh treatments of electrostatics.
We demonstrate the accuracy and efficiency of the qNEP approach through representative applications to water, \ce{Li7La3Zr2O12}, \ce{BaTiO3}, and a magnesium–water interface.
These results show that qNEP enables accurate atomistic simulations with explicit long-range electrostatics, scalable to million-atom systems on nanosecond time scales using consumer-grade GPUs.
\end{abstract}

\maketitle

\section{Introduction}

\Glspl{mlip} have become a widely adopted approach for accurate and efficient atomic-scale modeling of materials.
Early \glspl{mlip} \cite{behler2007generalized, bartok2010gaussian} were inherently short-ranged.
This approximation is adequate for many systems because of the short-sightedness of chemical bonding.
However, short-ranged models become inadequate in systems with sizable partial charges and weak screening, where electrostatic interactions are intrinsically long-ranged.
They are also limited when explicit coupling to external electric fields is required.

A common strategy to incorporate long-range electrostatics is to introduce fixed charges and subtract electrostatic contributions to energy and forces from the reference data \cite{AlbNorNor09, bartok2010gaussian, deng2019electrostatic}.
A more flexible alternative employs a separate regression model, such as a neural network, to predict partial charges, as in third-generation high-dimensional neural network potentials \cite{Artrith2011high, Morawietz2012neural}.
In this framework, partial charges are fitted to reference values obtained from a static charge decomposition scheme.
Such an approach is conceptually unsatisfactory because there is no unique decomposition of the electronic charge density into individual ionic contributions.
Other methods avoid explicit ionic charge partitioning by targeting higher-order electrostatic observables, such as the dipole moment \cite{unke2019physnet}, or by representing long-range electrostatics using the centers of maximally localized Wannier functions \cite{Zhang2022jcp, gao2022self}.

More recently, charge equilibration schemes originally developed for conventional interatomic potentials \cite{rappe1991charge, Ghasemi2015prb} have been adapted for use with \glspl{mlip} \cite{ko2021fourth, ko2023accurate}.
In contrast to approaches in which partial charges are predicted directly by a regression model, charge equilibration schemes determine the charges self-consistently by minimizing an electrostatic energy functional subject to global constraints.
This formulation enforces charge conservation and enables a physically consistent description of long-range charge transfer.
Such schemes are employed, for example, in the fourth-generation high-dimensional neural network potential \cite{ko2021fourth}, but they substantially increase the computational cost due to the expensive charge equilibration step, even when iterative solvers are used for acceleration \cite{Gubler2024jtct}.

An alternative route to charge conservation is to start from the electric enthalpy and obtain the \glspl{bec} as derivatives of the polarization.
While this approach is physically elegant and internally consistent, current implementations \cite{Falletta2025} rely on equivariant neural networks, which are computationally demanding.

To circumvent the reliance on reference charges and to at least partly alleviate the need for explicit charge equilibration schemes, several approaches have been developed in which partial charges are not learned explicitly.
Instead, they are treated as latent features of the model and determined implicitly by fitting the sum of the electrostatic energy and a short-range \gls{mlip} to the total target energies and forces.
Song \textit{et al.} \cite{song2024charge} treated the partial charges by including both real-space (short-ranged) and reciprocal-space (long-ranged) electrostatic contributions, whereas Cheng \textit{et al.} considered only the reciprocal-space (long-ranged) component \cite{cheng2025latent, King2025nc}.
The latter, so-called \gls{les} approach, also enables the calculation of the polarization and \glspl{bec} \cite{zhong2025machine}, and is available as a PyTorch-based library \cite{kim2025jctc}.
Related ideas have also been explored within a variational charge equilibration framework, which likewise enables the learning of partial charges without reference values \cite{shaidu2024incorporating}.

Although the general principles for the incorporation of electrostatics into \glspl{mlip} have been established, existing approaches remain computationally demanding.
As a result, their application to large-scale systems comprising hundreds of thousands to millions of atoms, as well as to long time scales extending from several to tens of nanoseconds, is severely constrained.
This limits their use in studies of phenomena that critically depend on long-range electrostatics and polarization, such as ion and proton transport, charged defects and defect migration, dielectric response and vibrational or infrared spectroscopy, and field-driven polarization dynamics.
Even for smaller systems, improving computational efficiency is essential to enable extensive sampling and an efficient use of modern computing resources.

In the present work, we therefore develop qNEP, a charge-aware \gls{mlip} that combines physical fidelity with high computational efficiency, enabling predictive simulations across broad classes of materials and extended length and time scales.
The qNEP framework builds on the \gls{nep} scheme, a highly efficient short-ranged \gls{mlip} architecture with demonstrated accuracy and performance across a wide range of materials and applications \cite{dong2024molecular, ying2025advances, xu2025mega}.
Following earlier work \cite{song2024charge, cheng2025latent}, we treat the partial charges as latent features of the model and obtain the \glspl{bec} as derivatives of the polarization.
Charge conservation is already strongly encouraged during training through an explicit regularization term.
This requires only a small numerical adjustment during simulations to ensure physically consistent electrostatics.
This formulation enables the direct computation of dielectric properties and infrared spectra, as well as a consistent coupling to external electric fields.

By implementing the \gls{pppm} method \cite{hockney1988computer} to evaluate electrostatic interactions during \gls{md} simulations, we obtain a highly performant approach that is only 1.5 to 3 times slower than equivalently trained \gls{nep} models, while offering both enhanced functionality and improved accuracy.
We demonstrate the accuracy and efficiency of qNEP through representative applications to water, \ce{Li7La3Zr2O12}, \ce{BaTiO3}, and a magnesium–water interface.
These results show that qNEP enables accurate atomistic simulations with explicit long-range electrostatics, scalable to million-atom systems on nanosecond time scales using consumer-grade GPUs.

\begin{figure*}[htb]
\centering
\includegraphics{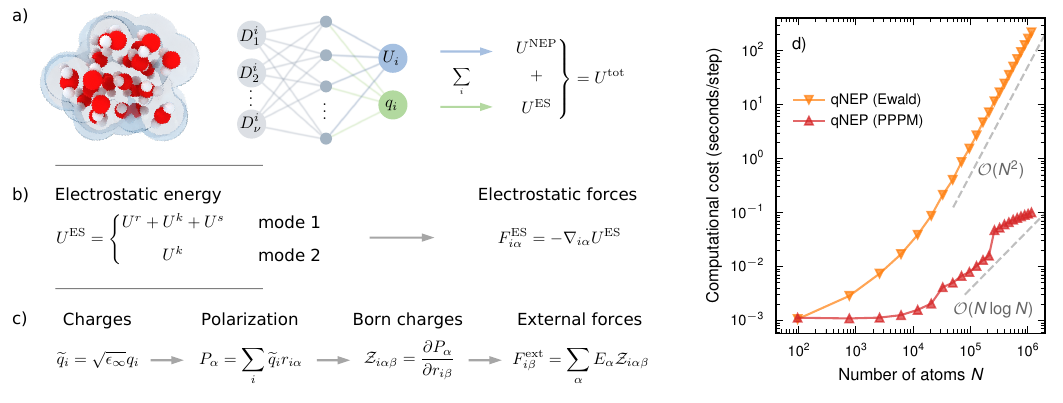}
\caption{
Schematic illustration of the qNEP framework. 
(a) Neural network architecture of a qNEP model with two outputs: a site energy $U_i$ and a partial charge $q_i$. 
The site energies are summed to yield the NEP contribution $U^\text{NEP}$ (\autoref{sect:nep}) to the total energy $U^\text{tot}$, while interactions between the partial charges give rise to the electrostatic energy $U^\text{ES}$ (\autoref{sect:qnep}).
(b) Evaluation of the electrostatic energy and corresponding forces $\vec{F}^\text{ES}_i$ (\autoref{sect:energy-derivatives}), including either both real- and reciprocal-space contributions (mode~1) or the reciprocal-space contribution only (mode~2).
(c) Derived response properties obtained from the partial charges, such as the polarization $\vec{P}$, the Born effective charges $\vec{\mathcal{Z}}_i$, and forces induced by external electric fields (\autoref{sect:born-effective-charges}).
(d) Computational cost for evaluating the electrostatic contribution, comparing direct Ewald summation with the \gls{pppm} method (\autoref{sect:ewald-and-pppm}), which offers superior computational performance, particularly for large systems (shown here for water; \autoref{sect:water}).
}
\label{fig:schematic}
\end{figure*}

\section{Methods \label{section:methods}}

\subsection{The original NEP model architecture}
\label{sect:nep}

The qNEP approach introduced below is based on the \gls{nep} framework \cite{fan2021neuroevolution}, which has undergone several refinements in recent years \cite{fan2022improving, fan2022gpumd, song2024general}.
In this section, we provide a brief overview of the most recent version, NEP4 \cite{song2024general}.
The term ``neuroevolution'' refers to the combination of a \gls{nn} model and an evolutionary training algorithm, namely the \gls{snes} \cite{schaul2011high}.

The machine-learning model used in \gls{nep} is a feed-forward \gls{nn} with a single hidden layer (\autoref{fig:schematic}a; blue output layer only).
In terms of the \gls{nn} model, the site energy can be explicitly expressed as
\begin{align}
    U_i = \sum_{\mu=1} ^{N_\mathrm{neu}} w ^{(1)} _{\mu} \tanh\left(\sum_{\nu=1} ^{N_\mathrm{des}} w ^{(0)}_{\mu\nu} D^i_{\nu} - b^{(0)}_{\mu}\right) - b^{(1)},
    \label{eq:U_i}
\end{align}
where $\tanh(x)$ is the activation function, $w^{(0)}$ represents the weight parameters connecting the input layer (with dimension $N_\text{des}$) and the hidden layer (with dimension $N_\text{neu}$), $w^{(1)}$ represents the weight parameters connecting the hidden layer and the output layer (the site energy), $b^{(0)}$ represents the bias parameters in the hidden layer, and $b^{(1)}$ represents the bias parameter in the output layer.
All of these parameters are trainable.

The total energy is given by the sum of the site energies
\begin{align}
    U^\text{NEP} = \sum_i U_i.
    \label{eq:U_NEP}
\end{align}

The input layer corresponds to the descriptor vector $\vec{D}^i$ (of dimension $N_\text{des}$) for a given atom $i$, with its components denoted as $D^i_{\nu}$ in \autoref{eq:U_i}.
Similar to the symmetry functions used in the Behler--Parrinello approach \cite{behler2007generalized, behler2011atom}, the descriptor components in \gls{nep} are classified into radial and angular ones.
Both types of descriptors involve additional trainable parameters used to discriminate between atomic species.
Details on the descriptor components and the associated trainable parameters can be found in Refs.~\citenum{xu2025mega, fan2022gpumd}.

\subsection{The qNEP model architecture}
\label{sect:qnep}

The qNEP approach extends the \gls{nep} framework by adding an additional output node to predict the partial charges $q_i$.
Using separate \glspl{nn} for the potential energy and the charges does not lead to a noticeable improvement in training accuracy, while increasing both the data requirements and the training cost.
We therefore adopt a single-\gls{nn} architecture (\autoref{fig:schematic}a; blue and green output layers).

Given the partial charges, the electrostatic energy $U^\text{ES}$ is evaluated under periodic boundary conditions using the Ewald decomposition as the sum of three contributions,
\begin{align}
    U^\text{ES} = U^\text{r} + U^{k} + U^\text{s},
    \label{eq:U_ES}
\end{align}
where $U^\text{r}$ is the real-space component, $U^{k}$ is the reciprocal-space component (evaluated in $k$-space, hence the superscript $k$), and $U^\text{s}$ is the self-energy.
The total energy in qNEP is given by the sum of the \gls{nep} energy and the electrostatic energy,
\begin{align}
    U^\text{tot} = U^\text{NEP} + U^\text{ES}.
    \label{eq:U_tot}
\end{align}

We consider two different modes for evaluating the energy contribution associated with the partial charges.
In mode~1, both the real-space and reciprocal-space contributions are included (\autoref{fig:schematic}b), as adopted by Song \textit{et al.} \cite{song2024charge}, and the electrostatic energy of the system is calculated according to \autoref{eq:U_ES}.
One may, however, argue that a short-ranged \gls{mlip} such as \gls{nep} already captures all short-range interactions, including the real-space component of electrostatics, making an explicit real-space electrostatic term potentially redundant.
In mode~2, we therefore consider only the reciprocal-space contribution (\autoref{fig:schematic}b), as adopted by Cheng \textit{et al.} \cite{cheng2025latent, King2025nc}, i.e.,
\begin{align}
U^\text{ES} = U^{k}.
\label{eq:mode2}
\end{align}
One of the aims of this work is to evaluate the relative advantages and disadvantages of these two approaches in realistic systems.

In the remainder of this section, we present explicit expressions for the real-space and reciprocal-space contributions to the electrostatic energy, as well as for the self-energy.
We then derive expressions for the force and virial (\autoref{sect:energy-derivatives}), as well as for the \gls{bec} tensor and related properties (\autoref{sect:born-effective-charges}).
The latter enables coupling to external electric fields and the computation of quantities such as the dielectric function, infrared spectra, and ionic electrical conductivity.

During training of qNEP models, the \gls{nep} loss function is augmented with additional terms that penalize violations of charge conservation and, optionally, constrain the prediction of the \glspl{bec} (\autoref{sect:loss-function}).
For the evaluation of reciprocal-space contributions, we implement both Ewald summation and the \gls{pppm} technique (\autoref{sect:ewald-and-pppm}).
We have made functionality for training and deploying qNEP models available in the free-and-open-source GPUMD package \cite{xu2025mega} from version 4.6, together with the supporting \textsc{calorine} Python package from version 3.3 \cite{LinRahFra24}.

\subsubsection{Real-space electrostatic energy}

The real-space electrostatic energy included in $U^\text{ES}$ when using mode~1 (\autoref{eq:U_ES}) is given by
\begin{align}
U^\text{r} = \frac{1}{2} \frac{1}{4\pi\epsilon_0} \sum_i \sum_{j\neq i}\frac{q_i q_j}{r_{ij}} \text{erfc}(\alpha , r_{ij}),
\end{align}
where $\text{erfc}$ denotes the complementary error function, $r_{ij}$ is the distance between atoms $i$ and $j$, $q_i$ is the charge of atom $i$, and $\epsilon_0$ is the vacuum permittivity.
We use the terms ``ion'' and ``atom'' interchangeably, as it is conventional to use atom in the context of \glspl{mlip}, while partial charges are typically associated with ions.
The real-space contribution is evaluated up to a cutoff radius $r_\text{c}$.

The parameter $\alpha$, which has the dimension of inverse length, controls the relative convergence rates of the real-space and reciprocal-space components.
Larger values of $\alpha$ lead to faster convergence in real space with respect to the cutoff radius $r_\text{c}$, whereas smaller values of $\alpha$ improve convergence in reciprocal space with respect to the cutoff wave vector $k_\text{max}$ (see below).
In this work, the real-space cutoff radius $r_\text{c}$ is chosen to coincide with the pairwise cutoff of the associated \gls{nep} model, which typically lies in the range \qtyrange{4}{8}{\angstrom}.
After fixing the \gls{nep} cutoff radius, we select $\alpha$ to provide sufficient accuracy for the real-space contribution.
Specifically, we use $\alpha=\pi/r_\text{c}$, which is a conventional choice to converge the real-space contribution \cite{toukmaji1996ewald}.

\subsubsection{Reciprocal-space electrostatic energy}

The reciprocal-space contribution to the electrostatic energy, $U^{k}$, which is required in both mode~1 (\autoref{eq:U_ES}) and mode~2 (\autoref{eq:mode2}), is given by
\begin{align}
\label{eq:U_k}
U^{k} = \frac{1}{4\pi\epsilon_0} \sum_{\vec{k}\neq \vec{0}}^{k<k_\text{max}} G(k) S(\vec{k}) S^{\ast}(\vec{k}),
\end{align}
where
\begin{align}
S(\vec{k}) \equiv \sum_i q_i e^{-\mathrm{i}\vec{k}\cdot\vec{r}_{i}} = S^{\ast}(-\vec{k})
\label{eq:S_k}
\end{align}
is the structure factor.
Here, $\vec{r}_i$ denotes the position of atom $i$, $\vec{k}$ is a reciprocal-space wave vector given by integer combinations of the reciprocal lattice basis vectors, and $k = |\vec{k}|$.

The function
\begin{align}
G(k) \equiv \frac{2\pi}{\Omega} \frac{1}{k^2} e^{-k^2/4\alpha^2},
\end{align}
where $\Omega$ is the volume of the simulation cell, corresponds to the product of the Green’s function of the Coulomb potential and a Gaussian smoothing function.
The summation over wave vectors $\vec{k}$ in \autoref{eq:U_k} excludes the $\vec{k}=\vec{0}$ term, which corresponds to the total charge of the system, and is truncated at a maximum magnitude $k_\text{max}$.
An accuracy of approximately \num{e-5}, consistent with that of the real-space contribution, is achieved by choosing $k_\text{max} = 2\pi\alpha$.

\subsubsection{Self-energy}

For mode~1 (\autoref{eq:U_ES}), we also include the self-energy term, which removes the unphysical interaction of each charge with its own screening cloud,
\begin{align}
U^\text{s} = -\frac{1}{4\pi\epsilon_0} \frac{\alpha}{\sqrt{\pi}} \sum_i q_i^2,
\end{align}
which is consistent with the approach adopted by Song \textit{et al.} \cite{song2024charge}.

\subsection{Energy derivatives}
\label{sect:energy-derivatives}

Starting from the energy, one can derive other microscopic quantities such as the force and virial.
A crucial aspect in the present context is that both static and dynamic contributions of the charges must be taken into account when evaluating energy derivatives.
Here, static charge refers to contributions originating from the explicit $1/r$ dependence in the Coulomb energy, whereas dynamic charge refers to contributions arising from the position dependence of the charges themselves.

qNEP models are many-body potentials, and general expressions for the force and virial of such potentials have been discussed previously \cite{fan2015force}.
An important result is that Newton’s third law (in its weak form) continues to hold for many-body potentials.
Accordingly, the force acting on atom $i$ can be expressed as a pairwise summation \cite{fan2015force}
\begin{align}
\vec{F}_i = \sum_{j \neq i} \left(\vec{F}_{ij}-\vec{F}_{ji}\right),
\end{align}
where $\vec{F}_{ij}$ can be interpreted as a ``partial force'' contribution.

Using the partial forces, the per-atom virial tensor can be written as \cite{fan2015force}
\begin{align}
\label{eq:per-atom_virial}
\vec{W}_i = \sum_{j \neq i} \vec{r}_{ij} \otimes \vec{F}_{ji}.
\end{align}
Throughout this work, we define
\begin{align}
\vec{r}_{ij} \equiv \vec{r}_j - \vec{r}_i
\end{align}
as the distance vector pointing from atom $i$ to atom $j$.

Analogous to the electrostatic energy the force is evaluated using an Ewald decomposition into real-space, reciprocal-space, and self-energy contributions.
For static charges, the real-space term gives rise to purely pairwise partial forces, while the reciprocal-space contribution is most naturally expressed as a per-atom force.
When the charges depend on the atomic configuration, additional force contributions arise in all three parts of the Ewald sum through the chain rule, i.e., from terms proportional to $(\partial E/\partial q_i)(\partial q_i/\partial \vec{r})$.
These dynamic-charge contributions can be cast into a partial-force form and combined consistently with the static terms, allowing the total force and virial to be evaluated within the same many-body framework.

We now present explicit expressions for the partial forces associated with the different energy contributions in the qNEP model.
For the \gls{nep} contribution, the partial force can be written as
\begin{align}
\vec{F}^\text{NEP}_{ij}
= \sum_{\nu=1}^{N_\text{des}} \frac{\partial U_i}{\partial D^i_{\nu}} \frac{\partial D^i_{\nu}}{\partial \vec{r}_{ij}}.
\end{align}
Details on the derivatives of the descriptors with respect to atomic positions can be found in previous work \cite{fan2022gpumd}.
For the electrostatic contribution, we discuss the three components separately in the following subsections.

For the contributions due to the dynamic charges the per-atom virial is obtained from the corresponding partial forces according to \autoref{eq:per-atom_virial}.

\subsubsection{The real-space contribution}

For static charges, the real-space contribution to the electrostatic energy is purely two-body (pairwise) in nature.
Nevertheless, it can be formulated within the general many-body potential framework introduced above.
Within this framework, the partial force can be derived as
\begin{align}
\vec{F}_{ij}^{\text{r, static}} = \frac{1}{2}\frac{-q_iq_j}{4\pi\epsilon_0} 
\frac{\vec{r}_{ij}}{r_{ij}^3} \left[\frac{2\alpha}{\sqrt{\pi}}r_{ij}e^{-\alpha^2r_{ij}^2}+\mathrm{erfc}(\alpha \, r_{ij}) \right].
\end{align}

When the charges are dynamic, i.e., explicitly dependent on the atomic configuration, an additional contribution to the partial force arises from the position dependence of the charges.
This contribution is given by
\begin{align}
\vec{F}_{ij}^{\text{r, dynamic}} = \frac{1}{4\pi\epsilon_0} 
\frac{\partial q_i}{\partial \vec{r}_{ij}} \left(\sum_{k\neq i} \frac{q_k}{r_{ik}} \mathrm{erfc}(\alpha \, r_{ik})\right).
\end{align}
The derivative of the charge with respect to the relative position vector is evaluated using the chain rule,
\begin{align} 
\frac{\partial q_i}{\partial \vec{r}_{ij}} = \sum_{\nu=1}^{N_\text{des}} \frac{\partial q_i}{\partial D^i_{\nu}}  \frac{\partial D^i_{\nu}}{\partial \vec{r}_{ij}}. 
\end{align}

\subsubsection{The reciprocal-space contribution}

The force acting on atom $i$ due to the reciprocal-space contribution of the electrostatic energy with static charges can be derived to be
\begin{align}
\vec{F}_i^{k, \text{static}} =  2\frac{q_i}{4\pi\epsilon_0} \sum_{\vec{k}\neq \vec{0}}^{k<k_\text{max}} \vec{k} \, G(k) \mathrm{Im} \left[S(\vec{k}) e^{\mathrm{i}\vec{k}\cdot\vec{r}_{i}}\right].
\end{align}

The partial force due to dynamic charges can be derived to be
\begin{align}
\vec{F}_{ij}^{k, \text{dynamic}}= 2\sum_{\vec{k}\neq \vec{0}}^{k<k_\text{max}} \frac{G(k)}{4\pi\epsilon_0}  \mathrm{Re}\left[S(\vec{k}) \frac{\partial q_i}{\partial \vec{r}_{ij}} e^{\mathrm{i}\vec{k}\cdot \vec{r}_i} \right].
\end{align}

While the reciprocal-space contribution to the force due to static charges is usually not calculated in a pair-wise manner, the virial can be calculated in a per-atom style \cite{Heyes1994prb},
\begin{align}
\vec{W}_i^{k, \text{static}} 
=  \sum_{\vec{k}\neq \vec{0}}^{k<k_\text{max}} \frac{G(k) \, q_i \, e^{\text{i}\vec{k} \cdot \vec{r}_i} \, S(\vec{k})}{4\pi\epsilon_0} \, \vec{B}.
\end{align}
Here, $\vec{B}$ is a $k$-space stress kernel that results from the derivative of the reciprocal-space electrostatic energy with respect to a homogeneous strain of the simulation cell and maps each $\vec{k}$-mode contribution onto a second-rank virial tensor,
\begin{align}
    \vec{B} = \vec{I} - \left(\frac{2}{4\alpha^2}+\frac{2}{k^2}\right)\vec{K},
\end{align}
where $\vec{I}$ is the $3 \times 3$ identity tensor, and $\vec{K}$ is a tensor with components $K_{\alpha\beta}=k_{\alpha}k_{\beta}$.
If the per-atom virial is not needed, the total virial can be more cheaply calculated as
\begin{align}
\label{eq:virial-kspace-total}
\vec{W}^{k, \text{static}} 
=  \sum_{\vec{k}\neq \vec{0}}^{k<k_\text{max}} \frac{G(k)|S(\vec{k})|^2}{4\pi\epsilon_0} 
\vec{B}.
\end{align}
Note that although $\vec{B}$ is a $3\times 3$ tensor, it has only six independent components, reflecting the symmetry of the stress tensor and the fact that the reciprocal-space contribution ultimately derives from an underlying pairwise electrostatic interaction.

\subsubsection{The self-energy contribution}

For static charges, the self-energy does not contribute to the force, since it depends only on the fixed charge values and is therefore independent of the atomic positions.
When the charges are configuration-dependent, however, the self-energy acquires an implicit position dependence through $q_i(\{\vec{r}\})$, which gives rise to an additional force contribution.
In this case, the corresponding partial force is
\begin{align}
\vec{F}_{ij}^{\text{s}, \text{dynamic}} = -\frac{2}{4\pi\epsilon_0} \frac{\alpha}{\sqrt{\pi}}  q_i \frac{\partial q_i}{\partial \vec{r}_{ij}}.
\end{align}
As in the other contributions, the derivative of the charge with respect to position is evaluated using the chain rule introduced above.

\subsection{Born effective charge and related properties}
\label{sect:born-effective-charges}

The output charges $q$ can be used to compute the macroscopic polarization $\vec{P}$ and the associated \glspl{bec} (\autoref{fig:schematic}c), as first discussed by Zhong \textit{et al.} \cite{zhong2025machine}
Before calculating the polarization, the learned partial charges need to be scaled \cite{Kirby2019jpcl}
\begin{align}
    \widetilde{q}_i = \sqrt{\epsilon_{\infty} }q_i, 
\end{align}
where $\epsilon_{\infty}$ denotes the high-frequency relative permittivity, also known as the electronic dielectric constant.
Using these notations, the Coulomb potential between two charges can be written as
\begin{align}
    \frac{1}{4\pi\epsilon_0} \frac{q_iq_j}{r_{ij}} 
    = \frac{1}{4\pi\epsilon_0\epsilon_{\infty}} \frac{\widetilde{q}_i\widetilde{q}_j}{r_{ij}}. 
\end{align}
This rescaling accounts for electronic screening effects that are not explicitly included in the ionic degrees of freedom.
The learned partial charges $q_i$ are thus screened charges, while the scaled charges $\widetilde{q}_i$ can be understood as naked charges.
The high-frequency relative permittivity $\epsilon_{\infty}$ is material specific and is generally taken as a trainable parameter.
In this work, we assume an isotropic dielectric constant, leaving an anisotropic extension for future work.

For non-periodic systems, where absolute positions are well defined, the polarization (which reduces to the dipole moment) can be written as
\begin{align}
    P_{\alpha} = \sum_i \widetilde{q}_i r_{i\alpha}
    \label{eq:polarization}
\end{align}
and the \gls{bec} tensors can be obtained as
\begin{align}
    \mathcal{Z}_{i\alpha\beta} 
    &= \frac{\partial P_{\alpha}}{\partial r_{i\beta}} = \widetilde{q}_i \delta_{\alpha\beta} + \sum_j^N r_{j\alpha} \frac{\partial \widetilde{q}_j}{\partial r_{i\beta}} \nonumber \\
    &= \widetilde{q}_i \delta_{\alpha\beta} 
    - \sum_{j\neq i} 
    \left(
    r_{i\alpha} \frac{\partial \widetilde{q}_i}{\partial r_{ij\beta}}
    -r_{j\alpha} \frac{\partial \widetilde{q}_j}{\partial r_{ji\beta}}
    \right),
    \label{eq:bec_1}
\end{align}
where the second expression follows from rewriting the derivatives in terms of relative position vectors.
For periodic systems, absolute positions are not well defined, and the polarization must be expressed in a translationally invariant form,
\begin{align}
\mathcal{Z}_{i\alpha\beta} = 
\widetilde{q}_i \delta_{\alpha\beta} 
    + \frac{1}{2}\sum_{j\neq i} 
    \left(
    r_{ij\alpha} \frac{\partial \widetilde{q}_i}{\partial r_{ij\beta}} -
    r_{ji\alpha} \frac{\partial \widetilde{q}_j}{\partial r_{ji\beta}}
    \right).
    \label{eq:bec_2}
\end{align}

Using the \gls{bec}, the force acting on ion $i$ in response to an external electric field $\vec{E}$ can be written as
\begin{align}
    F_{i\beta}^\text{ext}
    = \sum_{\alpha} E_{\alpha} \mathcal{Z}_{i\alpha\beta}.
    \label{eq:electric_force}
\end{align}

During \gls{md} simulations, the time derivative of the polarization corresponds to the ionic electric current and can be evaluated from the \gls{bec} and the atomic velocities $\vec{v}$ as
\begin{align}
    \dot{P}_{\alpha} = \frac{\text{d} P_{\alpha}}{\text{d} t} = \sum_{i=1}^N \sum_{\beta} \mathcal{Z}_{i\alpha\beta} v_{i\beta}.
    \label{eq:polarization_derivative}
\end{align}
The polarization along the trajectory can then be obtained by time integration of $\dot{P}_{\alpha}$, provided that the initial value is known.
The Fourier transform of the time \gls{acf} $\langle \dot{\vec{P}}(0) \cdot \dot{\vec{P}}(t) \rangle$ is proportional to the infrared spectrum, while its time integral yields the ionic electrical conductivity
\begin{align}
\sigma = \frac{1}{3k_\text{B}TV} \int_0^{\infty} \langle \dot{\vec{P}}(0) \cdot \dot{\vec{P}}(t) \rangle \text{d}t.
\end{align}

\subsection{Training of the models}
\label{sect:loss-function}

All parameters in the descriptor and the \gls{nn} for the potential energy and partial charges are trainable, including the high-frequency relative permittivity $\epsilon_{\infty}$.
As in the original \gls{nep} approach, these parameters are optimized using the \gls{snes} method \cite{schaul2011high}.
The optimization is guided by a loss function, which we denote as $L(\vec{z})$, where the abstract vector $\vec{z}$ collects all trainable parameters.

The loss function is defined as a weighted sum of the \gls{rmse} values for the energies ($\Delta_\text{e}$), forces ($\Delta_\text{f}$), virials ($\Delta_\text{v}$), \glspl{bec} ($\Delta_\mathcal{Z}$), and total charges ($\Delta_\text{Q}$), together with $\mathcal{L}_1$ and $\mathcal{L}_2$ regularization terms,
\begin{align}
L(\vec{z}) 
&= 
\lambda_\text{e} \Delta_\text{e}(\vec{z}) + 
\lambda_\text{f} \Delta_\text{f}(\vec{z}) + 
\lambda_\text{v} \Delta_\text{v}(\vec{z}) +
\lambda_\mathcal{Z} \Delta_\mathcal{Z}(\vec{z})
\nonumber \\
&+ 
\lambda_\text{Q} \Delta_\text{Q}(\vec{z}) + 
\lambda_1 \|\vec{z}\|_1 + \lambda_2\|\vec{z}\|_2^2.
\end{align}
Here, $\Delta_\text{Q}$ refers to the total charge of each structure rather than to individual partial charges and is included to penalize violations of charge conservation.
This penalty ensures that the predicted total charge deviates from the target value only marginally.

To enforce strict charge conservation or charge neutrality, a final total-charge correction is applied before evaluating the electrostatic energy and the \glspl{bec}.
Specifically, the scaled charges $\widetilde{q}_i$ in a system with $N$ atoms are corrected as follows,
\begin{equation}
    \widetilde{q}_i \rightarrow \widetilde{q}_i - \frac{1}{N}\sum_i (Q - \widetilde{q}_i),
    \label{eq:charge-scaling}
\end{equation}
where $Q$ is the target total charge of the structure, which is zero in all the cases studied in this work.
With the penalization term $\lambda_\text{Q} \Delta_\text{Q}(\vec{z})$ in the loss function, the total charge of the structure is already close to the target $Q$ and the correction above mainly serves to ensure strict charge conservation that can be important in, e.g., simulations with an external electric field.

The inclusion of target \glspl{bec} in the loss function is optional, and reference data need only be provided for a subset of the training structures.
This makes it possible to limit the number of reference \gls{bec} calculations, which are computationally more demanding than calculations of energies, forces, or virials.

\subsection{Accelerated calculation of the reciprocal-space contribution using PPPM}
\label{sect:ewald-and-pppm}

In the preceding sections, we assumed a direct Ewald summation for evaluating the reciprocal-space contribution to the electrostatic energy.
In practical simulations, however, the use of \gls{fft}-based methods can significantly reduce the computational cost \cite{Allen2017book}.
This leads to particle--mesh approaches such as \gls{pppm} \cite{hockney1988computer}, particle--mesh Ewald (PME) \cite{Darden1993jcp}, and smooth PME (SPME) \cite{Essmann1995jcp}, which are closely related and can be mathematically transformed into one another \cite{Deserno1998jcp, Ballenegger2012jctc}.
Here, we adopt the \gls{pppm} method and extend it to consistently account for both static and dynamic partial charges.

Within the \gls{pppm} framework, the reciprocal-space contribution to the electrostatic energy retains the formal structure of \autoref{eq:U_k}, but the Green’s function factor $G(k)$ is replaced by an optimized counterpart, $G^\text{opt}(\vec{k})$ \cite{Allen2017book},
\begin{align}
    U^{k} = \frac{1}{4\pi\epsilon_0} \sum_{\vec{k}\neq \vec{0}} G^\text{opt}(\vec{k}) S(\vec{k}) S^{\ast}(\vec{k}).
\end{align}
The structure factor $S(\vec{k})$ is evaluated on a regular mesh of dimension $N_x \times N_y \times N_z$.
Mesh charges are obtained by interpolating the original partial charges using a charge assignment function $W(\vec{r}_i - \vec{r}_s)$ \cite{Allen2017book}, which specifies the fraction of the charge at position $\vec{r}_i$ assigned to the mesh point $\vec{r}_s$.
The charge assignment function can be decomposed into the three Cartesian directions,
\begin{align}
    W(\vec{r}_i - \vec{r}_s) = W(x_i - x_s)
    \, W(y_i - y_s)
    \, W(z_i - z_s).
    \end{align}
Explicit expressions for the charge assignment functions for interpolation orders $P=1$ to $P=7$ are given by Deserno and Holm \cite{Deserno1998jcp}.

In our implementation, we use a mesh spacing smaller than \qty{1}{\angstrom} together with an interpolation order $P=5$, which yields an accuracy of approximately \num{e-4}.
The corresponding optimized Green’s function is given by \cite{Allen2017book, Ballenegger2012jctc}
\begin{align}
    G^\text{opt}(\vec{k}) =  
    \frac{G(k) \left[\prod_{\alpha=1}^3 \text{sinc}^P\left(\frac{\pi n_{\alpha}}{N_{\alpha}}\right)\right]^2}
    {\prod_{\alpha}^3 \left(1 - \frac{5}{3}z_{\alpha}^2 + \frac{7}{9}z_{\alpha}^4 + \frac{17}{189}z_{\alpha}^6 + \frac{2}{2835} z_{\alpha}^8\right)},
    \end{align}
where $\text{sinc}(x)=\sin(x)/x$, $z_{\alpha}=\sin\left(\pi n_{\alpha}/N_{\alpha}\right)$, and $n_{\alpha}$ are integer mesh indices satisfying $-N_{\alpha}/2 \leq n_{\alpha} < N_{\alpha}/2$.

The reciprocal-space force due to static charges can be computed in several equivalent ways \cite{Allen2017book}.
Here, we employ the $\mathrm{i}\vec{k}$-differentiation scheme,
\begin{align}
\label{eq:force-static-fft}
    \vec{F}_i^{k, \text{static}} = -\frac{2}{4\pi\epsilon_0}q_i \sum_{\vec{r}_s} W(\vec{r}_i - \vec{r}_s) \mathcal{F}^{-1} \left[\mathrm{i}\vec{k}S(\vec{k})G^\text{opt}(\vec{k})\right].
\end{align}
For dynamic charges, the force contribution arises from the explicit position dependence of the charges and is evaluated using an analytical differentiation scheme,
\begin{align}
\label{eq:force-dynamic-fft}
\begin{split}
    &\vec{F}_{ij}^{k, \text{dynamic}} \\
    &\quad= \frac{2}{4\pi\epsilon_0} \frac{\partial q_i}{\partial \vec{r}_{ij}}\sum_{\vec{r}_s}
    W(\vec{r}_i - \vec{r}_s) \, \mathcal{F}^{-1} \left[S(\vec{k})G^\text{opt}(\vec{k})\right].
\end{split}
\end{align}

For the reciprocal-space contribution to the virial arising from static charges, the total virial can be evaluated analogously to \autoref{eq:virial-kspace-total}.
If a per-atom decomposition is required, the per-atom virial can be written as
\begin{align}
\label{eq:per-atom-virial-fft}
    \vec{W}_i^{k, \text{static}} = \frac{q_i}{4\pi\epsilon_0} 
    \sum_{\vec{r}_s} W(\vec{r}_i - \vec{r}_s)
    \, \mathcal{F}^{-1} \left[ S(\vec{k})G^\text{opt}(\vec{k}) \vec{B}\right].
\end{align}

A forward \gls{fft} is used to compute $S(\vec{k})$ from the charge mesh, while backward \glspl{fft} are used to evaluate the forces and virials, as indicated by the $\mathcal{F}^{-1}$ operations.
Specifically, three backward \glspl{fft} are required to obtain the three Cartesian components of the force due to static charges according to \autoref{eq:force-static-fft}, and one backward \gls{fft} is required to obtain the force contribution due to dynamic charges according to \autoref{eq:force-dynamic-fft}.
If the per-atom virial is needed, six backward \glspl{fft} are required to evaluate the six independent components of the virial tensor according to \autoref{eq:per-atom-virial-fft}.
Efficient implementations of these operations can be realized using standard libraries from the CUDA and HIP toolkits.

The resulting \gls{pppm} implementation exhibits an overall $\mathcal{O}(N \log N)$ scaling with the number of atoms due to the use of \glspl{fft}, in contrast to the quadratic scaling of the direct Ewald summation.
In practice, the \gls{pppm} method features a small prefactor and near-linear scaling over the system sizes considered here, resulting in a computational cost that is one to several orders of magnitude lower than for the Ewald approach (\autoref{fig:schematic}d).

\begin{figure*}
\centering
\includegraphics{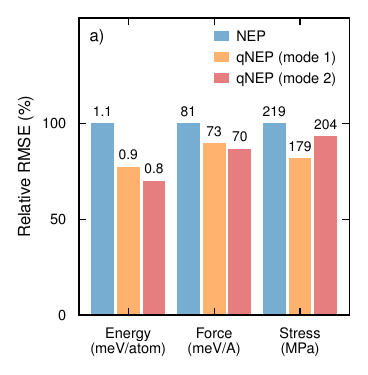}%
\includegraphics{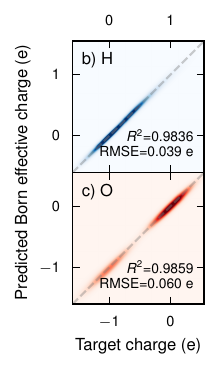}%
\includegraphics{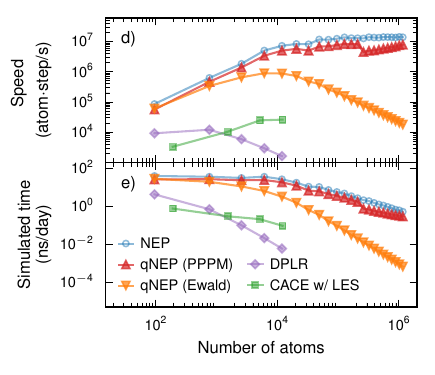}
\caption{
\textbf{Performance of qNEP for water.}
(a) Relative validation \glspl{rmse} of NEP and qNEP models.
(b,c) Parity plots of Born effective charges for H and O, respectively, obtained from the qNEP model trained on the full reference data set using mode~2.
(d,e) Performance comparison of qNEP models with \gls{nep}, a CACE model with \gls{les}, and a deep potential model with long-range electrostatic interactions (DPLR), in terms of (d) computational speed and (e) the simulated time achievable within one day on a single Nvidia RTX4090 GPU using a time step of \qty{0.5}{\femto\second}.
}
\label{fig:water-models}
\end{figure*}

\section{Results}
\label{section:results}
 
To illustrate the potential of the qNEP approach, we constructed models for several distinct classes of materials and employed them in prototypical applications.
In the following section (\autoref{sect:water}), we consider water as a representative liquid system.
We show that the inclusion of electrostatics in qNEP models systematically improves accuracy compared to regular \gls{nep} models at only a modest additional computational cost.
The resulting models enable simulations of water systems comprising hundreds of thousands or even millions of atoms and allow for simulation times of several tens of nanoseconds per day on a single GPU.
We further demonstrate the capability of qNEP models to predict the infrared spectrum of water as a function of temperature.

We then turn to two crystalline systems, again observing systematic improvements upon including electrostatics.
First, for the prototypical ionic conductor \ce{Li7La3Zr2O12}, we show that the temperature dependence of the structural parameters and the transition from the low-temperature tetragonal phase to the high-temperature cubic phase are in close agreement with experimental data (\autoref{sect:llzo}).
Further analysis reveals a qualitative change in the charge distribution across the phase transition, which is reflected in the ionic electrical conductivity, with the activation energy decreasing from \qty{1.45}{\electronvolt} in the tetragonal phase to \qty{0.29}{\electronvolt} in the cubic phase.

Next, we consider the prototypical ferroelectric \ce{BaTiO3}, demonstrating that qNEP models readily reproduce not only the experimentally observed phase transitions and structural changes, but also the associated evolution of the polarization (\autoref{sect:barium-titanate}).
We map out polarization--electric field (poling) loops at different temperatures, illustrating the coupling to external electric fields.
In addition, we extract the temperature dependence of both the dielectric function and the dielectric constant.

Finally, we examine magnesium corrosion in aqueous media, a reactive solid--liquid interface that combines metallic and insulating components (\autoref{sect:magnesium-corrosion}).
The qNEP approach captures the diverse, environment-dependent charge states present in this system and, owing to its computational efficiency, enables simulations of the conversion of metallic Mg into hydroxylated and solvated species under highly reactive conditions over timescales of many nanoseconds.

\begin{figure}
\centering
\includegraphics{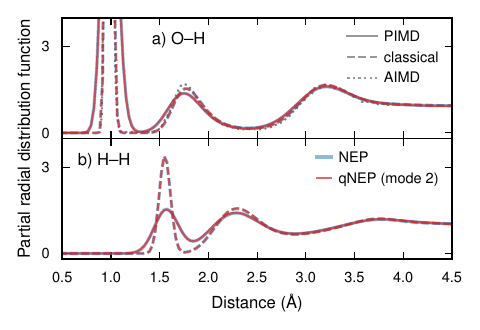}
\includegraphics{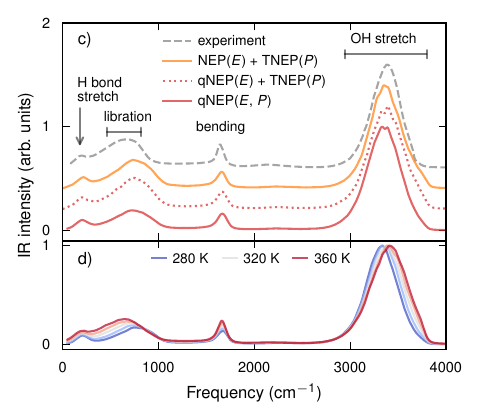}
\caption{
\textbf{Properties of water with NEP and qNEP models.}
(a,b) Partial radial distribution functions for (a) O--H and (b) H--H pairs in liquid water at \qty{300}{\kelvin}.
Solid and dashed lines correspond to quantum simulations performed using \gls{pimd} and classical simulations using standard \gls{md}, respectively, for the NEP (blue) and qNEP (red) models.
Classical ab-initio \gls{md} (AIMD) reference data (dotted line) from Ref.~\citenum{XuHaoLia23} are included for comparison.
(c,d) Infrared spectra obtained from classical \gls{md} simulations via the time \gls{acf} of the ionic electric current $\dot{\vec{P}}$.
(c) Spectra obtained using different combinations of the NEP and qNEP (mode~2) models for sampling the energy landscape ($E$) and the TNEP (from Ref.~\citenum{XuRosSch24}) and qNEP models for computing the polarization ($P$) in comparison with experiment.
(d) Temperature dependence of infrared spectra obtained using the qNEP model (mode~2) for both $E$ and $P$.
}
\label{fig:water-applications}
\end{figure}

\subsection{Liquid water}
\label{sect:water}

Water is a representative liquid system, in which the electrostatic between the components plays an important role.
To train models, we employed the data set of Zhang \textit{et al.} \cite{ZhaWanCar21} as curated by Xu \textit{et al.} \cite{XuHaoLia23}, who provided a split into \num{1388} training and \num{500} validation structures.
All structures contain \num{384} atoms, and energies, forces, and stresses were obtained from \gls{dft} calculations using the \gls{scan} exchange--correlation functional \cite{SunRuzPer15} (see Refs.~\citenum{ZhaWanCar21, XuHaoLia23} for details).
In addition, to enable learning of the dielectric response, we computed \glspl{bec} for \num{195} structures randomly selected from the original data set (see \autoref{snote:water-dft-calculations} for details) \cite{KreFur96a, Blo94, KreJou99}.
We trained one NEP model and two qNEP models (one for each electrostatic mode) using identical hyperparameters (\autoref{snote:water-training}, \autoref{sfig:water-parity-plots}).

The \glspl{rmse} demonstrate a systematic improvement in the accuracy of energies, forces, and stresses for the qNEP models compared to the \gls{nep} model (\autoref{fig:water-models}a), highlighting the importance of long-range electrostatic interactions in water.
The two qNEP variants perform very similarly with the model trained using only the reciprocal-space contribution (mode~2) yielding marginally lower errors for energies and forces.
This trend, which is also observed for the other systems discussed below, suggests that explicitly including short-ranged electrostatic interactions may be redundant when such interactions are already captured by the underlying short-ranged \gls{mlip}.
Both qNEP models accurately reproduce the \glspl{bec} (\autoref{fig:water-models}b,c; see \autoref{sfig:water-bec-comparison} for the mode~1 model).

The qNEP models also learn the square root of the high-frequency dielectric constant, $\sqrt{\epsilon_\infty}$, which appears in \autoref{eq:charge-scaling} and corresponds to the refractive index at optical frequencies, $n$.
Although $\sqrt{\epsilon_\infty}$ primarily acts as a hyperparameter during training, it is noteworthy that the fitted values, $\sqrt{\epsilon_\infty}=n=\num{1.77}$ and \num{1.53} for modes~1 and~2, respectively, are in reasonable agreement with the experimental value of \num{1.33} at ambient conditions \cite{refractive_index_water_1985, refractive_index_water_1998}.
Such a comparison is meaningful here because $\epsilon_\infty$ can be expected to be relatively homogeneous over the training domain.

Computational efficiency is critical for production simulations.
The \gls{nep} model achieves speeds exceeding \qty{e7}{\atom\step\per\second} on a single consumer-grade GPU (Nvidia RTX~4090) for systems containing at least \num{e4} atoms (\autoref{sfig:water-performance}).
Using a time step of \qty{0.5}{\femto\second}, this corresponds to up to \qty{40}{\nano\second} of \gls{md} simulation per day (\autoref{fig:water-models}e; see also \autoref{sfig:water-performance} for results on other GPUs).
When employing qNEP models together with the \gls{pppm} method, the computational cost increases only by about a factor of two (\autoref{fig:water-models}e and \autoref{sfig:water-speed-ratios}).
These numbers are several orders of magnitude higher than those achievable with, for example, a CACE model with \gls{les} \cite{cheng2025latent} or a deep potential model with long-range electrostatic interactions \cite{Zhang2022jcp}.

As an additional validation, we computed partial radial distribution functions from both classical and \gls{pimd} simulations (\autoref{fig:water-applications}a,b; \autoref{snote:water-simulations}) \cite{NIST_water, YinZhoSve25}.
The results are essentially indistinguishable between the NEP and qNEP models and are in very good agreement with ab initio \gls{md} simulations performed using the same exchange--correlation functional \cite{XuHaoLia23}.

Finally, the availability of \glspl{bec} combined with the high computational efficiency of qNEP models enables straightforward calculation of infrared spectra from the time \gls{acf} of the polarization or its time derivative \cite{XuRosSch24} (see \autoref{snote:water-simulations} for details).
At room temperature, the resulting infrared spectra compare well with experimental reference data \cite{DowWil75, MaxCha09}, with remaining deviations attributable to the underlying exchange--correlation functional (\autoref{fig:water-applications}c).
Upon increasing temperature, we observe a blueshift of the O--H stretching band and a redshift of the librational band (\autoref{fig:water-applications}d), both of which can be attributed to a weakening of intermolecular vibrational coupling.

\begin{figure*}
\centering
\includegraphics{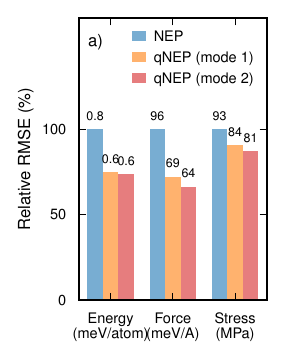}%
\includegraphics{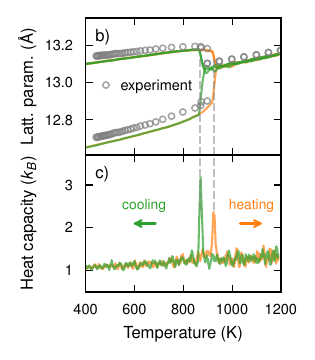}%
\includegraphics{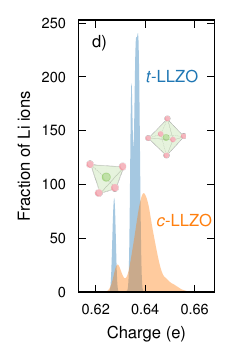}%
\includegraphics{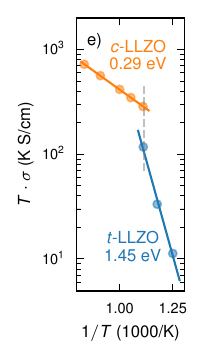}
\caption{
\textbf{Garnet-type \acrfull{llzo}.}
(a) Relative training \glspl{rmse} of \gls{nep} and qNEP models, with \gls{rmse} values reported above the corresponding columns.
(b,c) Temperature dependence of (b) lattice parameters and (c) heat capacity obtained from heating and cooling simulations.
Vertical dashed lines indicate the corresponding phase transition temperatures.
Experimental data for the lattice parameters from Ref.~\citenum{chen2015study}.
(d) Distribution of Li partial charges in the tetragonal (\textit{t}-\gls{llzo}) and cubic (\textit{c}-\gls{llzo}) phases after structural relaxation.
The left-hand peaks in the distributions correspond to Li-ions occupying tetrahedral sites (see inset) while the right-hand peaks correspond to octahedral sites (see inset).
(e) Arrhenius plot of the ionic conductivity $\sigma$ multiplied by temperature $T$.
Activation energies extracted for the tetragonal and cubic phases are indicated.
The vertical dashed line marks the average transition temperature of \qty{900}{\kelvin} obtained from heating and cooling runs.
}
\label{fig:llzo}
\end{figure*}

\subsection{Lithium lanthanum zirconate crystal}
\label{sect:llzo}

Garnet-type \gls{llzo} is among the most promising solid electrolyte materials for next-generation all-solid-state lithium batteries, combining high ionic conductivity with excellent chemical and electrochemical stability against lithium metal \cite{han2016electrochemical}.
Its garnet structure consists of a mobile Li-ion sublattice embedded within a rigid three-dimensional framework formed by interconnected \ce{LaO8} dodecahedra and \ce{ZrO6} octahedra, which creates fast diffusion pathways for Li-ions.
As an ionic crystalline solid electrolyte, \gls{llzo} represents an ideal test case for assessing the importance of incorporating charge information into \glspl{mlip}, since electrostatic interactions between charged species play a central role in ionic transport.
In addition, \gls{llzo} undergoes a well-known temperature-driven phase transition at approximately \qty{900}{\kelvin}, from a low-temperature tetragonal phase (\textit{t}-\gls{llzo}, $I4_1/acd$, ITC number~142) to a high-temperature cubic phase (\textit{c}-\gls{llzo}, $Ia\bar3d$, ITC number~230).
This transition is accompanied by an increase in the lithium ionic conductivity by several orders of magnitude \cite{yan2024impact} and has an order--disorder character, involving a reorganization of the Li-ion sublattice from an ordered arrangement with fully occupied sites in \textit{t}-\gls{llzo} to a disordered state with partially occupied sites in \textit{c}-\gls{llzo}.
The combination of strong electrostatic interactions and complex structural phase behavior makes \gls{llzo} a demanding and representative benchmark for evaluating the qNEP approach.

We trained NEP and qNEP models (\autoref{snote:llzo-training}) using the data set of Yan and Zhu \cite{yan2024impact}, which comprises \num{1978} configurations of pristine \gls{llzo} with energies, forces, and stresses computed using the PBEsol exchange--correlation functional \cite{Perdew2008_PBEsol}.
Consistent with the trends observed for liquid water, the qNEP models reduce the \glspl{rmse} for energies, forces, and stresses by approximately \qtyrange{20}{30}{\percent} relative to the regular \gls{nep} model (\autoref{fig:llzo}a; see also \autoref{sfig:llzo-parity-plots}).

Using the qNEP model trained in mode~2, we investigated the temperature dependence of the \gls{llzo} structure through heating and cooling simulations performed at a rate of \qty{50}{\kelvin\per\nano\second} (\autoref{snote:llzo-simulations}) \cite{MarTucTob96}.
The resulting lattice parameters, and in particular the thermal expansion, are in good agreement with experimental measurements over the full temperature range considered \cite{chen2015study} (\autoref{fig:llzo}b).
Our simulations capture the phase transition from \textit{t}-\gls{llzo} to \textit{c}-\gls{llzo} at approximately \qty{900}{\kelvin}, which is in excellent agreement with experimental observations. Additionally, a hysteresis of about \qty{55}{\kelvin} is observed between the heating and cooling cycles (\autoref{fig:llzo}c).

\begin{figure*}
\centering
\includegraphics{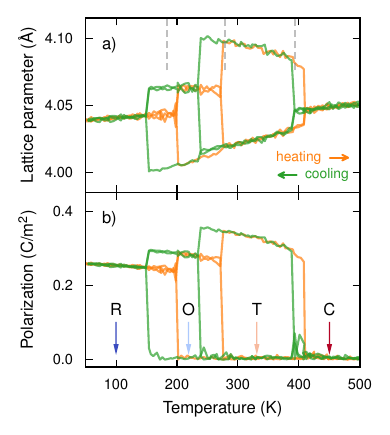}
\begin{minipage}[b]{2.1in}
\includegraphics{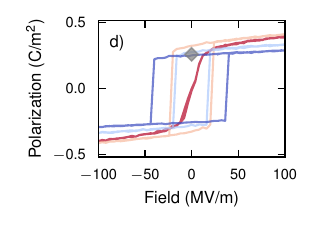}
\includegraphics{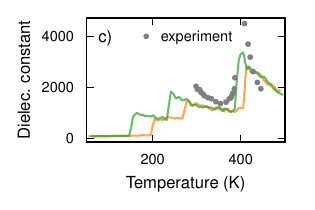}
\end{minipage}
\includegraphics{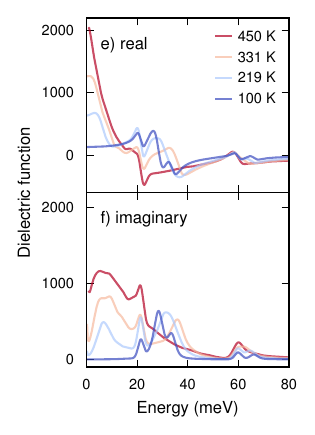}
\caption{
\textbf{Ferroelectricity and dielectric response in barium titanate (\ce{BaTiO3}).}
(a--c) Temperature dependence of (a) the lattice constant, (b) the polarization, and (c) the dielectric constant during heating and cooling, revealing the sequence of phase transitions from the rhombohedral (R) ground state through the orthorhombic (O) and tetragonal (T) phases to the high-temperature cubic (C) phase.
Vertical dashed lines indicate the experimentally observed transition temperatures \cite{Merz1949}.
Experimental data points in (c) from Ref.~\citenum{Benedict1958}.
(d) Polarization--electric field ($P$--$E$) hysteresis loops and (e,f) the real and imaginary parts of the dielectric function at different temperatures, corresponding to all four phases (indicated by arrows in (b)).
The gray diamond in (d) marks the experimental value for the spontaneous polarization at room temperature \cite{Wieder1955}.
}
\label{fig:bto-switching}
\end{figure*}

The environment-dependent dynamic charges predicted by the qNEP model enable a detailed analysis of the order--disorder transition in \gls{llzo}.
We relaxed snapshots extracted from \gls{md} trajectories of both \textit{t}-\gls{llzo} and \textit{c}-\gls{llzo} and evaluated the distributions of Li-ion charges in each phase (\autoref{fig:llzo}d).
In \textit{t}-\gls{llzo}, the charge distribution exhibits a lower-charge peak associated with Li-ions occupying tetrahedral sites (Wyckoff position 8\textit{a}) and a higher-charge peak corresponding to ions in octahedral sites (16\textit{f} and 32\textit{g}).
The octahedral contribution further displays a split structure, which we attribute to the distinct occupations of the 16\textit{f} and 32\textit{g} Wyckoff sites, both of which are octahedrally coordinated but feature slightly different local environments.
In \textit{c}-\gls{llzo}, the charge distribution becomes broader, reflecting the disordered nature of the Li-ion sublattice in the cubic phase.

These structural and charge-distribution differences between \textit{t}-\gls{llzo} and \textit{c}-\gls{llzo} are directly reflected in the transport properties (\autoref{fig:llzo}e).
In particular, the ionic conductivity exhibits a pronounced reduction in activation energy, decreasing from \qty{1.45}{\electronvolt} in \textit{t}-\gls{llzo} to \qty{0.29}{\electronvolt} in \textit{c}-\gls{llzo}.
The latter value is in good agreement with experimental data in the high-temperature region \cite{WanLai15}.
The activation energy in \textit{t}-\gls{llzo} is known to be highly sensitive to composition \cite{yan2024impact}, whence a direct comparison is not possible but a possible target for future work.

\begin{figure}
\centering
\includegraphics{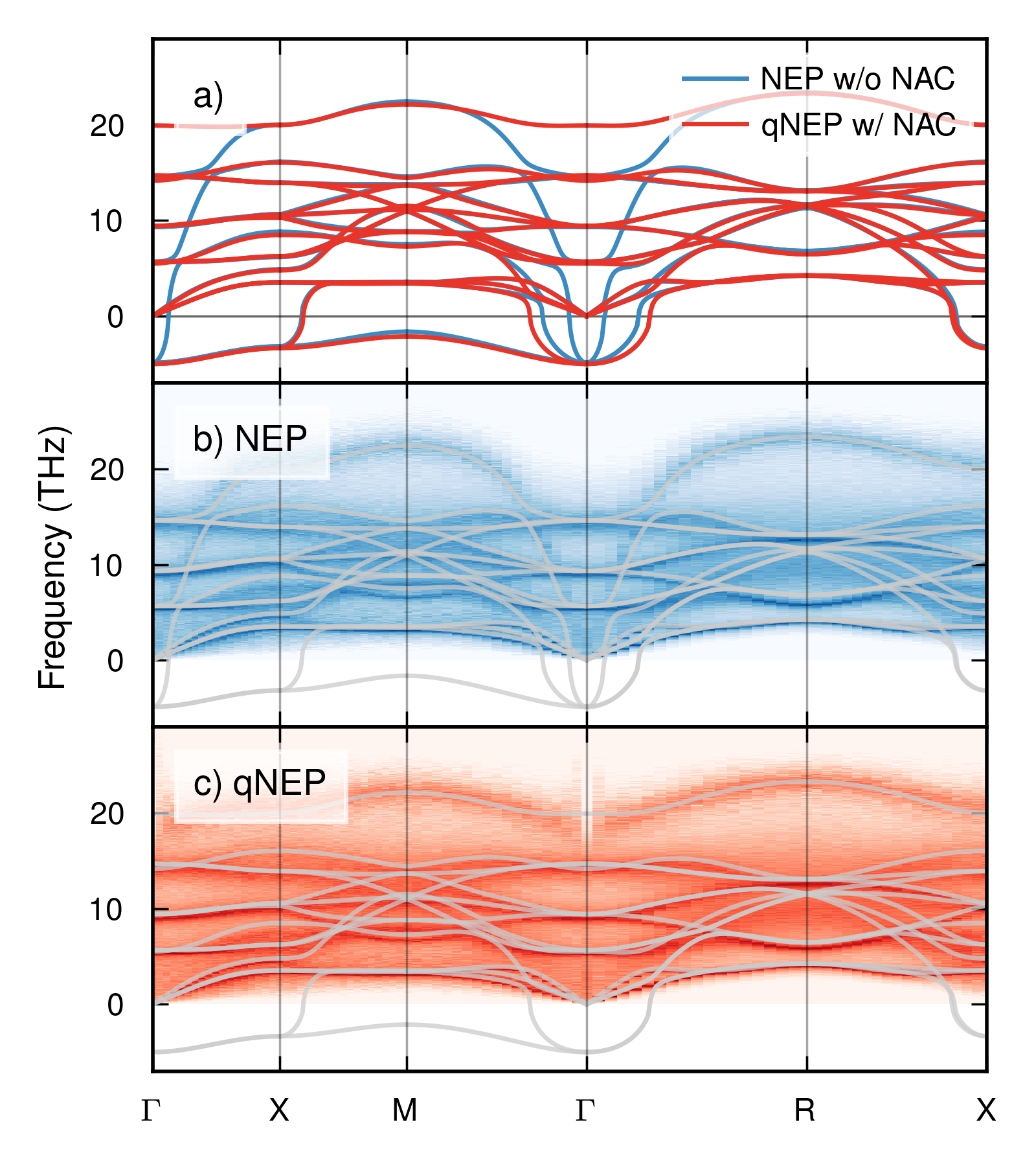}
\caption{
\textbf{Phonons and LO--TO splitting in \ce{BaTiO3}.}
(a) Harmonic phonon dispersion of the cubic phase obtained from \gls{nep} without \acrfull{nac} and from qNEP including \acrshort{nac}, illustrating the longitudinal--transverse optical (LO--TO) splitting.
Imaginary phonon frequencies are shown as negative values and indicate dynamically unstable modes.
(b,c) Spectral energy density obtained from (b) \gls{nep} and (c) qNEP at \qty{500}{\kelvin} in the cubic-phase region calculated with \textsc{dynasor} \cite{dynasor1, dynasor2}.
Gray lines reproduce the corresponding harmonic phonon dispersions shown in (a).
}
\label{fig:bto-phonons}
\end{figure}

\subsection{Barium titanate}
\label{sect:barium-titanate}

Next, we consider the prototypical ferroelectric perovskite \ce{BaTiO3}, which provides an ideal test case for evaluating qNEP models in the presence of strong electromechanical coupling and external electric fields.
At low temperatures, \ce{BaTiO3} adopts a rhombohedral structure with an instantaneous polarization along the $\langle 111\rangle$ direction due to off-centering of the Ti atoms \cite{Merz1949, Gigli2022}.
Upon heating, it undergoes a sequence of phase transitions to orthorhombic and tetragonal phases at \qty{183}{\kelvin} and \qty{278}{\kelvin}, respectively, with polarization along the $\langle 011\rangle$ and $\langle 001\rangle$ directions \cite{Merz1949}.
At temperatures above \qty{393}{\kelvin}, the material becomes paraelectric and cubic.

\Gls{nep} and qNEP models were trained using a data set extended from Lindgren \textit{et al.} \cite{LinJacFra25} to \num{1832} structures, \num{1193} of which also included \glspl{bec} (\autoref{snote:bto-training}).
Energies, forces, and stresses were obtained from \gls{dft} calculations \cite{KreFur96a} using the r2SCAN exchange--correlation functional \cite{Furness2020}, while \glspl{bec} were obtained using the PBEsol functional \cite{Perdew2008_PBEsol} (\autoref{snote:bto-dft-calculations}).
As in previous cases, the qNEP models achieve a higher accuracy than the corresponding NEP model.
The qNEP model based on mode~1 and used for the simulations below achieves \glspl{rmse} of \qty{1.2}{\milli\electronvolt\per\atom}, \qty{65}{\milli\electronvolt\per\angstrom}, and \qty{112}{\giga\pascal} for energies, forces, and stresses (\autoref{sfig:bto-parity-plots}).
For comparison, the corresponding \glspl{rmse} for the NEP model are \qty{1.0}{\milli\electronvolt\per\atom}, \qty{70}{\milli\electronvolt\per\angstrom}, and \qty{136}{\giga\pascal}.
Both qNEP models also perform very well at predicting the \glspl{bec} (\autoref{sfig:bto-bec-comparison}).

The trained model correctly reproduces the sequence of four phases and yields transition temperatures of \qty{151}{\kelvin}, \qty{235}{\kelvin}, and \qty{390}{\kelvin}, in good agreement with experiment (\autoref{fig:bto-switching}a; \autoref{snote:bto-simulations}).
This agreement can be attributed to the accuracy of the underlying exchange--correlation functional, whose energetics are faithfully reproduced by the qNEP model.
A hysteresis of up to \qty{50}{\kelvin} is observed between heating and cooling runs, even at the comparatively low rate of \qty{10}{\kelvin\per\nano\second} used here.
This behavior reflects the first-order nature of the phase transitions, despite their relatively small latent heats.

Using the predicted \glspl{bec}, we directly computed the polarization as a function of temperature (\autoref{fig:bto-switching}b).
All four phases are clearly resolved, with a pronounced polarization at low temperatures and a vanishing polarization above \qty{407}{\kelvin} during heating and \qty{390}{\kelvin} during cooling.
The polarization increases when progressing from the rhombohedral through the orthorhombic to the tetragonal phase, in agreement with both shell-model simulations \cite{Hashimoto2015} and experimental measurements \cite{Merz1949}.

For selected temperatures representative of each phase, we further applied external electric fields to map out polarization--electric field ($P$--$E$) hysteresis loops (\autoref{fig:bto-switching}d; \autoref{snote:bto-simulations}).
As expected, the polarization can be switched by the applied field, and the spontaneous polarization in the absence of a field agrees well with the experimental value at room temperature \cite{Wieder1955} (indicated by the gray diamond in \autoref{fig:bto-switching}d).
The coercive field depends sensitively on the switching frequency.
Here, a frequency of \qty{500}{\mega\hertz} was employed, which is low for \gls{md} simulations but still significantly higher than experimentally accessible frequencies \cite{Jiang2022}.
A direct quantitative comparison of coercive fields is therefore not meaningful.
Nevertheless, the computational efficiency of the qNEP approach opens the door to detailed investigations of switching mechanisms and domain-wall motion, particularly within multiscale simulation frameworks \cite{ShiGriChe07}.

As a further validation, we computed the ionic contribution to the dielectric function from the time \gls{acf} of the ionic electric current (\autoref{snote:bto-simulations}).
The results reveal a strong dependence of the dielectric response on both temperature and frequency (\autoref{fig:bto-switching}e,f).
Resonances appearing in the range \qtyrange{20}{40}{\milli\electronvolt} can be attributed to longitudinal optical modes at the $\Gamma$ point.
Their pronounced temperature dependence reflects the strong anharmonicity of \ce{BaTiO3}, which also gives rise to the very large dielectric constant observed (\autoref{fig:bto-switching}c; see also the static limit of the real part in \autoref{fig:bto-switching}e).
The maximum dielectric constant obtained here reaches values of approximately \num{3000} on the high-temperature side of the tetragonal--cubic phase boundary, and is in good agreement with experimental measurements both in terms of magnitude and temperature dependence \cite{Benedict1958}.
In this context, it is important to emphasize that the present simulations only include the ionic (vibrational) contribution while the experimental measurements also contain contributions from processes such as space-charge effects and grain boundaries.
These occur on much longer time scales and their relative importance depends on sample preparation.

Finally, we examine the effect of long-range electrostatics on the phonon dispersion in the cubic phase.
The harmonic phonon dispersions, calculated with \textsc{phonopy} \cite{togo2023implementation, phonopy2023}, predicted by the NEP and qNEP models agree closely over most of the Brillouin zone, except in the vicinity of the $\Gamma$ point (\autoref{fig:bto-phonons}a; \autoref{snote:bto-phonons}).
This difference arises from long-range Coulomb interactions, which lead to a splitting between longitudinal and transverse optical phonon branches.
Accounting for this LO--TO splitting requires the application of \glspl{nac}, which in turn depend on knowledge of the \glspl{bec} and are therefore only accessible within the qNEP framework.
This distinction becomes even more pronounced in finite-temperature phonon dispersions obtained from the spectral energy density \cite{Thomas2010} using dynasor \cite{dynasor1, dynasor2} (\autoref{fig:bto-phonons}b,c), where the LO and TO modes coincide at $\Gamma$ for the NEP model but remain clearly separated for qNEP.

\begin{figure*}
\centering
\includegraphics{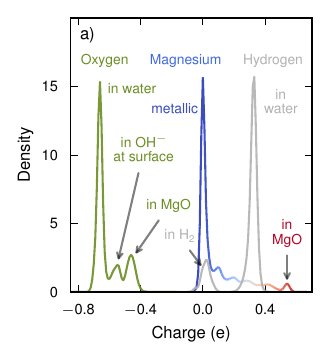}%
\includegraphics{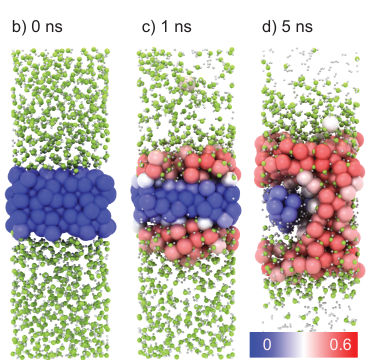}%
\includegraphics{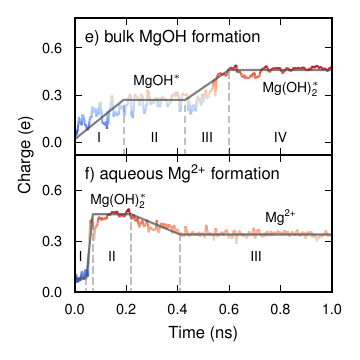}
\caption{
\textbf{Corrosion of magnesium at a Mg--water interface.}
(a) Distribution of atomic charges in the reference data set, distinguishing Mg, O, and H species across different chemical environments.
(b--d) Representative snapshots from an \gls{md} simulation illustrating the conversion of metallic Mg (large blue spheres) into oxidized Mg species (large red spheres).
The color scale (shown in (d)) indicates the Mg charge state.
Oxygen and hydrogen atoms are shown as small green and gray spheres, respectively.
(e,f) Two characteristic charge-evolution pathways of Mg: (e) formation of bulk magnesium hydroxide and (f) dissolution into aqueous \ce{Mg^{2+}} ions.
Asterisks ($*$) denote adsorbed species.
}
\label{fig:MgOH}
\end{figure*}

\subsection{Corrosion of magnesium in water}
\label{sect:magnesium-corrosion}

Magnesium corrosion in aqueous media provides a prototypical example of a chemically reactive system in which accurate treatment of charge transfer is essential.
Changes in the Mg valence state directly govern reaction pathways, intermediate species, and final corrosion products.
In particular, the widely discussed incomplete-film monovalent magnesium-ion mechanism invokes the presence of a transient \ce{Mg+} intermediate and remains controversial \cite{BenGomCos22}.
Resolving this issue is central to understanding the anomalous hydrogen evolution observed during magnesium corrosion and to guiding the design of corrosion-resistant magnesium alloys.

In previous studies, standard \glspl{mlip} were unable to explicitly represent evolving charge states and their associated long-range electrostatic interactions.
This limitation hindered detailed investigations of charge evolution during the formation of corrosion intermediates and the dissolution dynamics of magnesium.
The magnesium--water solid--liquid interface therefore constitutes a stringent and representative test case for qNEP models, as it directly probes their ability to capture environment-dependent charge transfer in reactive chemical processes.

To train NEP and qNEP models, we employed the reference data set of Liu \textit{et al.} \cite{liu2025cej}, which systematically covers magnesium dissolution mechanisms (see \autoref{snote:MgOH-training} for details).
The qNEP model achieves \glspl{rmse} values of \qty{12}{\milli\electronvolt\per\atom} for energies, \qty{198}{\milli\electronvolt\per\angstrom} for forces, and \qty{824}{\giga\pascal} for stresses, representing a clear improvement over the NEP model, which yields \qty{14}{\milli\electronvolt\per\atom}, \qty{230}{\milli\electronvolt\per\angstrom}, and \qty{787}{\giga\pascal}, respectively (\autoref{sfig:MgOH-parity-plots}).
Crucially, the qNEP model successfully learns the relationship between local atomic environments and charge distributions (\autoref{fig:MgOH}a).
The predicted charges clearly distinguish different chemical states: Mg atoms in the metallic bulk exhibit charges close to zero, while Mg atoms in hydroxides or oxides carry positive charges.
Oxygen atoms associated with water molecules, surface hydroxyl groups, and oxides likewise show distinct charge signatures.

\Gls{md} simulations using the qNEP model therefore cannot only capture the structural evolution of the Mg--water interface, but also provide direct insight into transitions between different charge states.
As a representative example, we consider the evolution of a highly reactive stepped Mg surface \cite{LiuBaoSha23} in contact with water at \qty{700}{\kelvin} over a simulation time of \qty{5}{\nano\second} that was recently analyzed using a \gls{nep} model (\autoref{fig:MgOH}b--d).
The detailed trajectories for such reactive systems are determined by a series of rare events, leading to considerable variety between individual simulations.
Here, rather than repeating the comprehensive analysis from Ref.~\citenum{liu2025cej}, we therefore applied our qNEP model to analyze a specific trajectory from this earlier work (see \autoref{snote:MgOH-simulations} for details).

After \qty{1}{\nano\second} (\autoref{fig:MgOH}c), the system clearly separates into an upper hydroxide layer (red), a lower metallic Mg substrate (blue), and a transitional interfacial region containing partially hydroxylated Mg species (light gray/pink).
In addition, a small number of Mg atoms undergo oxidation via dissolution into the aqueous phase.
As corrosion proceeds, most of the metallic Mg is converted into hydroxylated species within \qty{5}{\nano\second} (\autoref{fig:MgOH}d).
While Mg atoms within the bulk metal and the corrosion product layer retain relatively stable valence states, atoms in the interfacial region exhibit distinct intermediate charge states corresponding to partial hydroxylation.

A detailed analysis of the charge trajectories reveals two characteristic corrosion pathways.
The first pathway corresponds to the formation of solid-state corrosion products (\autoref{fig:MgOH}e).
Here, Mg atoms at the interface gradually increase their charge (stage~I), followed by stabilization at an intermediate value associated with \ce{MgOH{*}} species (stage~II; the asterisk ($*$) denotes an adsorbed species).
During this stage, near-surface Mg atoms are hydroxylated, whereas atoms deeper in the substrate remain metallic.
As \ce{OH{*}} species migrate further inward (stage~III), Mg undergoes deeper hydroxylation, forming amorphous \ce{Mg(OH)2{*}}, which can subsequently reorganize into crystalline \ce{Mg(OH)2}.
Under conditions favoring the formation of a protective surface film, this pathway dominates.

The second pathway corresponds to the dissolution of hydrated \ce{Mg^{2+}} ions into the aqueous phase (\autoref{fig:MgOH}f).
While the initial hydroxylation stages are similar to those of the solid-state pathway, Mg atoms located closer to the surface rapidly reach a divalent charge state (stage~II, labeled \ce{Mg(OH)2{*}} in \autoref{fig:MgOH}f).
These Mg species then detach from the surface, leaving hydroxyl groups behind on the substrate, and form solvated \ce{Mg^{2+}} ions in solution.

The qNEP model explicitly captures the competing mechanisms of solid-state oxidation and ionic dissolution, highlighting its ability to describe complex electrochemical interfaces with evolving charge states.
Overall, the application of qNEP to the Mg--water system not only reproduces key corrosion mechanisms identified in previous studies, but also provides dynamic, atom-resolved charge information that reveals the coupled evolution of valence states and structure during magnesium corrosion.
This example demonstrates that qNEP overcomes key limitations of traditional \glspl{mlip} and enables efficient, large-scale simulations of charge-transfer-driven processes such as corrosion and electrocatalysis.

\section{Summary and conclusions}

We have introduced qNEP, a charge-aware extension of the highly efficient \gls{nep} framework that incorporates explicit long-range electrostatics while retaining the computational performance required for large-scale \gls{md} simulations.
In qNEP, partial charges are learned as latent model features without relying on reference charge partitioning, charge conservation is enforced through a dedicated regularization term and a final total-charge correction, and polarization and \glspl{bec} follow consistently as derivatives of the learned charges.
By combining this formulation with an efficient \gls{pppm} implementation for reciprocal-space electrostatics, qNEP attains a computational cost only about 1.5 to 3 times higher than comparably trained \gls{nep} models, enabling simulations that extend to million-atom systems and nanosecond-to-tens-of-nanoseconds time scales on consumer-grade GPUs per day.
Models that include the short-ranged electrostatic contribution explicitly (mode~1) and those that include only the long-ranged contribution (mode~2) yield similar accuracy.

Across representative liquid, ionic, ferroelectric, and reactive-interface systems, qNEP systematically improves the accuracy of energies, forces, and stresses relative to \gls{nep} while providing direct access to charge- and field-related observables.
For water, qNEP delivers improved errors at modest overhead and enables infrared spectroscopy via the time \gls{acf} of the ionic electric current.
For garnet-type \ce{Li7La3Zr2O12}, qNEP captures the tetragonal--cubic transition and reveals phase-dependent charge distributions that correlate with the change in ionic transport barriers.
For \ce{BaTiO3}, qNEP reproduces the sequence of ferroelectric phase transitions and enables predictions of polarization dynamics, dielectric response, and LO--TO splitting through \glspl{nac} based on the predicted \glspl{bec}.
For magnesium corrosion in aqueous media, qNEP resolves environment-dependent charge states at a reactive solid--liquid interface and captures competing pathways of solid-state hydroxide formation and dissolution into aqueous \ce{Mg^{2+}}.
Together, these results establish qNEP as a practical and scalable route to accurate atomistic simulations of charge-transfer- and polarization-driven phenomena, opening the door to predictive studies of, e.g., transport, dielectric response, spectroscopy, and electrochemical reactivity across extended length and time scales.

Finally, we emphasize that the present work focuses exclusively on electrostatic interactions as the long-range contribution.
Extensions of the qNEP framework to incorporate other long-range interactions, such as dispersion forces \cite{yu2022capturing, Loche2025jcp, ji2025prl}, constitute a natural and promising direction for future research.
Such developments would further broaden the applicability of the \gls{nep} framework, in particular for aqueous, molecular, and biomolecular systems, where weak electronic screening in water renders electrostatic and other long-range interactions essential.

\section*{Supporting information}

Additional technical information and figures (PDF).

\section*{Data availability}

The \gls{nep} and qNEP models as well as the reference data used for their training and validation have been deposited on Zenodo under accession code \href{https://doi.org/10.5281/zenodo.18335947}{10.5281/zenodo.18335947}.

\section*{Acknowledgments}

ZF, BT, KX, and HD were supported by the Advanced Material National Science and Technology Major Project (grant No. 2025ZD0618902). 
EB, EB, EF, JW, and PE acknowledge funding from the Swedish Research Council (Nos.~2020-04935 and 2025-03999), the Knut and Alice Wallenberg Foundation (Nos.~2023.0032 and 2024.0042), the European Research Council (ERC Starting Grant No. 101162195), the Swedish Energy Agency (grant No. 45410-1), and the Swedish Strategic Research Foundation through a Future Research Leader programme (FFL21-0129).
ZY and YZ acknowledge support from the National Natural Science Foundation of China (Nos. 22509162 and 225B2917). 
ZL and ZW acknowledge support from the Taishan Scholars Youth Expert Program of Shandong Province (tsqn202312002).
ZS and LL acknowledge the Center for Computational Science and Engineering of the Southern University of Science and Technology.
The computations were enabled by resources provided by the National Academic Infrastructure for Supercomputing in Sweden (NAISS) at C3SE, PDC, and NSC, partially funded by the Swedish Research Council through grant agreement no. 2022-06725, the Berzelius resource provided by the Knut and Alice Wallenberg Foundation at NSC, as well as the Open Source Super-computing Center of S-A-I.


\begin{thebibliography}{80}%
\makeatletter
\providecommand \@ifxundefined [1]{%
 \@ifx{#1\undefined}
}%
\providecommand \@ifnum [1]{%
 \ifnum #1\expandafter \@firstoftwo
 \else \expandafter \@secondoftwo
 \fi
}%
\providecommand \@ifx [1]{%
 \ifx #1\expandafter \@firstoftwo
 \else \expandafter \@secondoftwo
 \fi
}%
\providecommand \natexlab [1]{#1}%
\providecommand \enquote  [1]{``#1''}%
\providecommand \bibnamefont  [1]{#1}%
\providecommand \bibfnamefont [1]{#1}%
\providecommand \citenamefont [1]{#1}%
\providecommand \href@noop [0]{\@secondoftwo}%
\providecommand \href [0]{\begingroup \@sanitize@url \@href}%
\providecommand \@href[1]{\@@startlink{#1}\@@href}%
\providecommand \@@href[1]{\endgroup#1\@@endlink}%
\providecommand \@sanitize@url [0]{\catcode `\\12\catcode `\$12\catcode `\&12\catcode `\#12\catcode `\^12\catcode `\_12\catcode `\%12\relax}%
\providecommand \@@startlink[1]{}%
\providecommand \@@endlink[0]{}%
\providecommand \url  [0]{\begingroup\@sanitize@url \@url }%
\providecommand \@url [1]{\endgroup\@href {#1}{\urlprefix }}%
\providecommand \urlprefix  [0]{URL }%
\providecommand \Eprint [0]{\href }%
\providecommand \doibase [0]{https://doi.org/}%
\providecommand \selectlanguage [0]{\@gobble}%
\providecommand \bibinfo  [0]{\@secondoftwo}%
\providecommand \bibfield  [0]{\@secondoftwo}%
\providecommand \translation [1]{[#1]}%
\providecommand \BibitemOpen [0]{}%
\providecommand \bibitemStop [0]{}%
\providecommand \bibitemNoStop [0]{.\EOS\space}%
\providecommand \EOS [0]{\spacefactor3000\relax}%
\providecommand \BibitemShut  [1]{\csname bibitem#1\endcsname}%
\let\auto@bib@innerbib\@empty
\bibitem [{\citenamefont {Behler}\ and\ \citenamefont {Parrinello}(2007)}]{behler2007generalized}%
  \BibitemOpen
  \bibfield  {author} {\bibinfo {author} {\bibfnamefont {J.}~\bibnamefont {Behler}}\ and\ \bibinfo {author} {\bibfnamefont {M.}~\bibnamefont {Parrinello}},\ }\bibfield  {title} {\bibinfo {title} {Generalized neural-network representation of high-dimensional potential-energy surfaces},\ }\href {https://doi.org/10.1103/PhysRevLett.98.146401} {\bibfield  {journal} {\bibinfo  {journal} {Physical Review Letters}\ }\textbf {\bibinfo {volume} {98}},\ \bibinfo {pages} {146401} (\bibinfo {year} {2007})}\BibitemShut {NoStop}%
\bibitem [{\citenamefont {Bart\'ok}\ \emph {et~al.}(2010)\citenamefont {Bart\'ok}, \citenamefont {Payne}, \citenamefont {Kondor},\ and\ \citenamefont {Cs\'anyi}}]{bartok2010gaussian}%
  \BibitemOpen
  \bibfield  {author} {\bibinfo {author} {\bibfnamefont {A.~P.}\ \bibnamefont {Bart\'ok}}, \bibinfo {author} {\bibfnamefont {M.~C.}\ \bibnamefont {Payne}}, \bibinfo {author} {\bibfnamefont {R.}~\bibnamefont {Kondor}},\ and\ \bibinfo {author} {\bibfnamefont {G.}~\bibnamefont {Cs\'anyi}},\ }\bibfield  {title} {\bibinfo {title} {Gaussian approximation potentials: The accuracy of quantum mechanics, without the electrons},\ }\href {https://doi.org/10.1103/PhysRevLett.104.136403} {\bibfield  {journal} {\bibinfo  {journal} {Physical Review Letters}\ }\textbf {\bibinfo {volume} {104}},\ \bibinfo {pages} {136403} (\bibinfo {year} {2010})}\BibitemShut {NoStop}%
\bibitem [{\citenamefont {Albe}\ \emph {et~al.}(2009)\citenamefont {Albe}, \citenamefont {Nord},\ and\ \citenamefont {Nordlund}}]{AlbNorNor09}%
  \BibitemOpen
  \bibfield  {author} {\bibinfo {author} {\bibfnamefont {K.}~\bibnamefont {Albe}}, \bibinfo {author} {\bibfnamefont {J.}~\bibnamefont {Nord}},\ and\ \bibinfo {author} {\bibfnamefont {K.}~\bibnamefont {Nordlund}},\ }\bibfield  {title} {\bibinfo {title} {Dynamic charge-transfer bond-order potential for gallium nitride},\ }\href {https://doi.org/10.1080/14786430903313708} {\bibfield  {journal} {\bibinfo  {journal} {Philosophical Magazine}\ }\textbf {\bibinfo {volume} {89}},\ \bibinfo {pages} {3477} (\bibinfo {year} {2009})}\BibitemShut {NoStop}%
\bibitem [{\citenamefont {Deng}\ \emph {et~al.}(2019)\citenamefont {Deng}, \citenamefont {Chen}, \citenamefont {Li},\ and\ \citenamefont {Ong}}]{deng2019electrostatic}%
  \BibitemOpen
  \bibfield  {author} {\bibinfo {author} {\bibfnamefont {Z.}~\bibnamefont {Deng}}, \bibinfo {author} {\bibfnamefont {C.}~\bibnamefont {Chen}}, \bibinfo {author} {\bibfnamefont {X.-G.}\ \bibnamefont {Li}},\ and\ \bibinfo {author} {\bibfnamefont {S.~P.}\ \bibnamefont {Ong}},\ }\bibfield  {title} {\bibinfo {title} {An electrostatic spectral neighbor analysis potential for lithium nitride},\ }\href {https://doi.org/10.1038/s41524-019-0212-1} {\bibfield  {journal} {\bibinfo  {journal} {npj Computational Materials}\ }\textbf {\bibinfo {volume} {5}},\ \bibinfo {pages} {75} (\bibinfo {year} {2019})}\BibitemShut {NoStop}%
\bibitem [{\citenamefont {Artrith}\ \emph {et~al.}(2011)\citenamefont {Artrith}, \citenamefont {Morawietz},\ and\ \citenamefont {Behler}}]{Artrith2011high}%
  \BibitemOpen
  \bibfield  {author} {\bibinfo {author} {\bibfnamefont {N.}~\bibnamefont {Artrith}}, \bibinfo {author} {\bibfnamefont {T.}~\bibnamefont {Morawietz}},\ and\ \bibinfo {author} {\bibfnamefont {J.}~\bibnamefont {Behler}},\ }\bibfield  {title} {\bibinfo {title} {High-dimensional neural-network potentials for multicomponent systems: Applications to zinc oxide},\ }\href {https://doi.org/10.1103/PhysRevB.83.153101} {\bibfield  {journal} {\bibinfo  {journal} {Physical Review B}\ }\textbf {\bibinfo {volume} {83}},\ \bibinfo {pages} {153101} (\bibinfo {year} {2011})}\BibitemShut {NoStop}%
\bibitem [{\citenamefont {Morawietz}\ \emph {et~al.}(2012)\citenamefont {Morawietz}, \citenamefont {Sharma},\ and\ \citenamefont {Behler}}]{Morawietz2012neural}%
  \BibitemOpen
  \bibfield  {author} {\bibinfo {author} {\bibfnamefont {T.}~\bibnamefont {Morawietz}}, \bibinfo {author} {\bibfnamefont {V.}~\bibnamefont {Sharma}},\ and\ \bibinfo {author} {\bibfnamefont {J.}~\bibnamefont {Behler}},\ }\bibfield  {title} {\bibinfo {title} {A neural network potential-energy surface for the water dimer based on environment-dependent atomic energies and charges},\ }\href {https://doi.org/10.1063/1.3682557} {\bibfield  {journal} {\bibinfo  {journal} {The Journal of Chemical Physics}\ }\textbf {\bibinfo {volume} {136}},\ \bibinfo {pages} {064103} (\bibinfo {year} {2012})}\BibitemShut {NoStop}%
\bibitem [{\citenamefont {Unke}\ and\ \citenamefont {Meuwly}(2019)}]{unke2019physnet}%
  \BibitemOpen
  \bibfield  {author} {\bibinfo {author} {\bibfnamefont {O.~T.}\ \bibnamefont {Unke}}\ and\ \bibinfo {author} {\bibfnamefont {M.}~\bibnamefont {Meuwly}},\ }\bibfield  {title} {\bibinfo {title} {Physnet: A neural network for predicting energies, forces, dipole moments, and partial charges},\ }\href {https://doi.org/10.1021/acs.jctc.9b00181} {\bibfield  {journal} {\bibinfo  {journal} {Journal of Chemical Theory and Computation}\ }\textbf {\bibinfo {volume} {15}},\ \bibinfo {pages} {3678} (\bibinfo {year} {2019})}\BibitemShut {NoStop}%
\bibitem [{\citenamefont {Zhang}\ \emph {et~al.}(2022)\citenamefont {Zhang}, \citenamefont {Wang}, \citenamefont {Muniz}, \citenamefont {Panagiotopoulos}, \citenamefont {Car},\ and\ \citenamefont {E}}]{Zhang2022jcp}%
  \BibitemOpen
  \bibfield  {author} {\bibinfo {author} {\bibfnamefont {L.}~\bibnamefont {Zhang}}, \bibinfo {author} {\bibfnamefont {H.}~\bibnamefont {Wang}}, \bibinfo {author} {\bibfnamefont {M.~C.}\ \bibnamefont {Muniz}}, \bibinfo {author} {\bibfnamefont {A.~Z.}\ \bibnamefont {Panagiotopoulos}}, \bibinfo {author} {\bibfnamefont {R.}~\bibnamefont {Car}},\ and\ \bibinfo {author} {\bibfnamefont {W.}~\bibnamefont {E}},\ }\bibfield  {title} {\bibinfo {title} {A deep potential model with long-range electrostatic interactions},\ }\href {https://doi.org/10.1063/5.0083669} {\bibfield  {journal} {\bibinfo  {journal} {The Journal of Chemical Physics}\ }\textbf {\bibinfo {volume} {156}},\ \bibinfo {pages} {124107} (\bibinfo {year} {2022})}\BibitemShut {NoStop}%
\bibitem [{\citenamefont {Gao}\ and\ \citenamefont {Remsing}(2022)}]{gao2022self}%
  \BibitemOpen
  \bibfield  {author} {\bibinfo {author} {\bibfnamefont {A.}~\bibnamefont {Gao}}\ and\ \bibinfo {author} {\bibfnamefont {R.~C.}\ \bibnamefont {Remsing}},\ }\bibfield  {title} {\bibinfo {title} {Self-consistent determination of long-range electrostatics in neural network potentials},\ }\href {https://doi.org/10.1038/s41467-022-29243-2} {\bibfield  {journal} {\bibinfo  {journal} {Nature Communications}\ }\textbf {\bibinfo {volume} {13}},\ \bibinfo {pages} {1572} (\bibinfo {year} {2022})}\BibitemShut {NoStop}%
\bibitem [{\citenamefont {Rappe}\ and\ \citenamefont {Goddard}(1991)}]{rappe1991charge}%
  \BibitemOpen
  \bibfield  {author} {\bibinfo {author} {\bibfnamefont {A.~K.}\ \bibnamefont {Rappe}}\ and\ \bibinfo {author} {\bibfnamefont {W.~A.~I.}\ \bibnamefont {Goddard}},\ }\bibfield  {title} {\bibinfo {title} {Charge equilibration for molecular dynamics simulations},\ }\href {https://doi.org/10.1021/j100161a070} {\bibfield  {journal} {\bibinfo  {journal} {The Journal of Physical Chemistry}\ }\textbf {\bibinfo {volume} {95}},\ \bibinfo {pages} {3358} (\bibinfo {year} {1991})}\BibitemShut {NoStop}%
\bibitem [{\citenamefont {Ghasemi}\ \emph {et~al.}(2015)\citenamefont {Ghasemi}, \citenamefont {Hofstetter}, \citenamefont {Saha},\ and\ \citenamefont {Goedecker}}]{Ghasemi2015prb}%
  \BibitemOpen
  \bibfield  {author} {\bibinfo {author} {\bibfnamefont {S.~A.}\ \bibnamefont {Ghasemi}}, \bibinfo {author} {\bibfnamefont {A.}~\bibnamefont {Hofstetter}}, \bibinfo {author} {\bibfnamefont {S.}~\bibnamefont {Saha}},\ and\ \bibinfo {author} {\bibfnamefont {S.}~\bibnamefont {Goedecker}},\ }\bibfield  {title} {\bibinfo {title} {Interatomic potentials for ionic systems with density functional accuracy based on charge densities obtained by a neural network},\ }\href {https://doi.org/10.1103/PhysRevB.92.045131} {\bibfield  {journal} {\bibinfo  {journal} {Physical Review B}\ }\textbf {\bibinfo {volume} {92}},\ \bibinfo {pages} {045131} (\bibinfo {year} {2015})}\BibitemShut {NoStop}%
\bibitem [{\citenamefont {Ko}\ \emph {et~al.}(2021)\citenamefont {Ko}, \citenamefont {Finkler}, \citenamefont {Goedecker},\ and\ \citenamefont {Behler}}]{ko2021fourth}%
  \BibitemOpen
  \bibfield  {author} {\bibinfo {author} {\bibfnamefont {T.~W.}\ \bibnamefont {Ko}}, \bibinfo {author} {\bibfnamefont {J.~A.}\ \bibnamefont {Finkler}}, \bibinfo {author} {\bibfnamefont {S.}~\bibnamefont {Goedecker}},\ and\ \bibinfo {author} {\bibfnamefont {J.}~\bibnamefont {Behler}},\ }\bibfield  {title} {\bibinfo {title} {A fourth-generation high-dimensional neural network potential with accurate electrostatics including non-local charge transfer},\ }\href {https://doi.org/10.1038/s41467-020-20427-2} {\bibfield  {journal} {\bibinfo  {journal} {Nature Communications}\ }\textbf {\bibinfo {volume} {12}},\ \bibinfo {pages} {398} (\bibinfo {year} {2021})}\BibitemShut {NoStop}%
\bibitem [{\citenamefont {Ko}\ \emph {et~al.}(2023)\citenamefont {Ko}, \citenamefont {Finkler}, \citenamefont {Goedecker},\ and\ \citenamefont {Behler}}]{ko2023accurate}%
  \BibitemOpen
  \bibfield  {author} {\bibinfo {author} {\bibfnamefont {T.~W.}\ \bibnamefont {Ko}}, \bibinfo {author} {\bibfnamefont {J.~A.}\ \bibnamefont {Finkler}}, \bibinfo {author} {\bibfnamefont {S.}~\bibnamefont {Goedecker}},\ and\ \bibinfo {author} {\bibfnamefont {J.}~\bibnamefont {Behler}},\ }\bibfield  {title} {\bibinfo {title} {Accurate fourth-generation machine learning potentials by electrostatic embedding},\ }\href {https://doi.org/10.1021/acs.jctc.2c01146} {\bibfield  {journal} {\bibinfo  {journal} {Journal of Chemical Theory and Computation}\ }\textbf {\bibinfo {volume} {19}},\ \bibinfo {pages} {3567} (\bibinfo {year} {2023})}\BibitemShut {NoStop}%
\bibitem [{\citenamefont {Gubler}\ \emph {et~al.}(2024)\citenamefont {Gubler}, \citenamefont {Finkler}, \citenamefont {Schäfer}, \citenamefont {Behler},\ and\ \citenamefont {Goedecker}}]{Gubler2024jtct}%
  \BibitemOpen
  \bibfield  {author} {\bibinfo {author} {\bibfnamefont {M.}~\bibnamefont {Gubler}}, \bibinfo {author} {\bibfnamefont {J.~A.}\ \bibnamefont {Finkler}}, \bibinfo {author} {\bibfnamefont {M.~R.}\ \bibnamefont {Schäfer}}, \bibinfo {author} {\bibfnamefont {J.}~\bibnamefont {Behler}},\ and\ \bibinfo {author} {\bibfnamefont {S.}~\bibnamefont {Goedecker}},\ }\bibfield  {title} {\bibinfo {title} {Accelerating fourth-generation machine learning potentials using quasi-linear scaling particle mesh charge equilibration},\ }\href {https://doi.org/10.1021/acs.jctc.4c00334} {\bibfield  {journal} {\bibinfo  {journal} {Journal of Chemical Theory and Computation}\ }\textbf {\bibinfo {volume} {20}},\ \bibinfo {pages} {7264} (\bibinfo {year} {2024})}\BibitemShut {NoStop}%
\bibitem [{\citenamefont {Falletta}\ \emph {et~al.}(2025)\citenamefont {Falletta}, \citenamefont {Cepellotti}, \citenamefont {Johansson}, \citenamefont {Tan}, \citenamefont {Descoteaux}, \citenamefont {Musaelian}, \citenamefont {Owen},\ and\ \citenamefont {Kozinsky}}]{Falletta2025}%
  \BibitemOpen
  \bibfield  {author} {\bibinfo {author} {\bibfnamefont {S.}~\bibnamefont {Falletta}}, \bibinfo {author} {\bibfnamefont {A.}~\bibnamefont {Cepellotti}}, \bibinfo {author} {\bibfnamefont {A.}~\bibnamefont {Johansson}}, \bibinfo {author} {\bibfnamefont {C.~W.}\ \bibnamefont {Tan}}, \bibinfo {author} {\bibfnamefont {M.~L.}\ \bibnamefont {Descoteaux}}, \bibinfo {author} {\bibfnamefont {A.}~\bibnamefont {Musaelian}}, \bibinfo {author} {\bibfnamefont {C.~J.}\ \bibnamefont {Owen}},\ and\ \bibinfo {author} {\bibfnamefont {B.}~\bibnamefont {Kozinsky}},\ }\bibfield  {title} {\bibinfo {title} {Unified differentiable learning of electric response},\ }\href {https://doi.org/10.1038/s41467-025-59304-1} {\bibfield  {journal} {\bibinfo  {journal} {Nature Communications}\ }\textbf {\bibinfo {volume} {16}},\ \bibinfo {pages} {4031} (\bibinfo {year} {2025})}\BibitemShut {NoStop}%
\bibitem [{\citenamefont {Song}\ \emph {et~al.}(2024{\natexlab{a}})\citenamefont {Song}, \citenamefont {Han}, \citenamefont {Henkelman},\ and\ \citenamefont {Li}}]{song2024charge}%
  \BibitemOpen
  \bibfield  {author} {\bibinfo {author} {\bibfnamefont {Z.}~\bibnamefont {Song}}, \bibinfo {author} {\bibfnamefont {J.}~\bibnamefont {Han}}, \bibinfo {author} {\bibfnamefont {G.}~\bibnamefont {Henkelman}},\ and\ \bibinfo {author} {\bibfnamefont {L.}~\bibnamefont {Li}},\ }\bibfield  {title} {\bibinfo {title} {Charge-optimized electrostatic interaction atom-centered neural network algorithm},\ }\href {https://doi.org/10.1021/acs.jctc.3c01254} {\bibfield  {journal} {\bibinfo  {journal} {Journal of Chemical Theory and Computation}\ }\textbf {\bibinfo {volume} {20}},\ \bibinfo {pages} {2088} (\bibinfo {year} {2024}{\natexlab{a}})}\BibitemShut {NoStop}%
\bibitem [{\citenamefont {Cheng}(2025)}]{cheng2025latent}%
  \BibitemOpen
  \bibfield  {author} {\bibinfo {author} {\bibfnamefont {B.}~\bibnamefont {Cheng}},\ }\bibfield  {title} {\bibinfo {title} {Latent ewald summation for machine learning of long-range interactions},\ }\href {https://doi.org/10.1038/s41524-025-01577-7} {\bibfield  {journal} {\bibinfo  {journal} {npj Computational Materials}\ }\textbf {\bibinfo {volume} {11}},\ \bibinfo {pages} {80} (\bibinfo {year} {2025})}\BibitemShut {NoStop}%
\bibitem [{\citenamefont {King}\ \emph {et~al.}(2025)\citenamefont {King}, \citenamefont {Kim}, \citenamefont {Zhong},\ and\ \citenamefont {Cheng}}]{King2025nc}%
  \BibitemOpen
  \bibfield  {author} {\bibinfo {author} {\bibfnamefont {D.~S.}\ \bibnamefont {King}}, \bibinfo {author} {\bibfnamefont {D.}~\bibnamefont {Kim}}, \bibinfo {author} {\bibfnamefont {P.}~\bibnamefont {Zhong}},\ and\ \bibinfo {author} {\bibfnamefont {B.}~\bibnamefont {Cheng}},\ }\bibfield  {title} {\bibinfo {title} {Machine learning of charges and long-range interactions from energies and forces},\ }\href {https://doi.org/10.1038/s41467-025-63852-x} {\bibfield  {journal} {\bibinfo  {journal} {Nature Communications}\ }\textbf {\bibinfo {volume} {16}},\ \bibinfo {pages} {8763} (\bibinfo {year} {2025})}\BibitemShut {NoStop}%
\bibitem [{\citenamefont {Zhong}\ \emph {et~al.}(2025)\citenamefont {Zhong}, \citenamefont {Kim}, \citenamefont {King},\ and\ \citenamefont {Cheng}}]{zhong2025machine}%
  \BibitemOpen
  \bibfield  {author} {\bibinfo {author} {\bibfnamefont {P.}~\bibnamefont {Zhong}}, \bibinfo {author} {\bibfnamefont {D.}~\bibnamefont {Kim}}, \bibinfo {author} {\bibfnamefont {D.~S.}\ \bibnamefont {King}},\ and\ \bibinfo {author} {\bibfnamefont {B.}~\bibnamefont {Cheng}},\ }\href {https://arxiv.org/abs/2504.05169} {\bibinfo {title} {Machine learning interatomic potential can infer electrical response}} (\bibinfo {year} {2025}),\ \Eprint {https://arxiv.org/abs/2504.05169} {arXiv:2504.05169 [cond-mat.mtrl-sci]} \BibitemShut {NoStop}%
\bibitem [{\citenamefont {Kim}\ \emph {et~al.}(2025)\citenamefont {Kim}, \citenamefont {Wang}, \citenamefont {Vargas}, \citenamefont {Zhong}, \citenamefont {King}, \citenamefont {Inizan},\ and\ \citenamefont {Cheng}}]{kim2025jctc}%
  \BibitemOpen
  \bibfield  {author} {\bibinfo {author} {\bibfnamefont {D.}~\bibnamefont {Kim}}, \bibinfo {author} {\bibfnamefont {X.}~\bibnamefont {Wang}}, \bibinfo {author} {\bibfnamefont {S.}~\bibnamefont {Vargas}}, \bibinfo {author} {\bibfnamefont {P.}~\bibnamefont {Zhong}}, \bibinfo {author} {\bibfnamefont {D.~S.}\ \bibnamefont {King}}, \bibinfo {author} {\bibfnamefont {T.~J.}\ \bibnamefont {Inizan}},\ and\ \bibinfo {author} {\bibfnamefont {B.}~\bibnamefont {Cheng}},\ }\bibfield  {title} {\bibinfo {title} {A universal augmentation framework for long-range electrostatics in machine learning interatomic potentials},\ }\href {https://doi.org/10.1021/acs.jctc.5c01400} {\bibfield  {journal} {\bibinfo  {journal} {Journal of Chemical Theory and Computation}\ }\textbf {\bibinfo {volume} {21}},\ \bibinfo {pages} {12709} (\bibinfo {year} {2025})}\BibitemShut {NoStop}%
\bibitem [{\citenamefont {Shaidu}\ \emph {et~al.}(2024)\citenamefont {Shaidu}, \citenamefont {Pellegrini}, \citenamefont {K{\"u}{\c{c}}{\"u}kbenli}, \citenamefont {Lot},\ and\ \citenamefont {de~Gironcoli}}]{shaidu2024incorporating}%
  \BibitemOpen
  \bibfield  {author} {\bibinfo {author} {\bibfnamefont {Y.}~\bibnamefont {Shaidu}}, \bibinfo {author} {\bibfnamefont {F.}~\bibnamefont {Pellegrini}}, \bibinfo {author} {\bibfnamefont {E.}~\bibnamefont {K{\"u}{\c{c}}{\"u}kbenli}}, \bibinfo {author} {\bibfnamefont {R.}~\bibnamefont {Lot}},\ and\ \bibinfo {author} {\bibfnamefont {S.}~\bibnamefont {de~Gironcoli}},\ }\bibfield  {title} {\bibinfo {title} {Incorporating long-range electrostatics in neural network potentials via variational charge equilibration from shortsighted ingredients},\ }\href {https://doi.org/10.1038/s41524-024-01225-6} {\bibfield  {journal} {\bibinfo  {journal} {npj Computational Materials}\ }\textbf {\bibinfo {volume} {10}},\ \bibinfo {pages} {47} (\bibinfo {year} {2024})}\BibitemShut {NoStop}%
\bibitem [{\citenamefont {Dong}\ \emph {et~al.}(2024)\citenamefont {Dong}, \citenamefont {Shi}, \citenamefont {Ying}, \citenamefont {Xu}, \citenamefont {Liang}, \citenamefont {Wang}, \citenamefont {Zeng}, \citenamefont {Wu}, \citenamefont {Zhou}, \citenamefont {Xiong}, \citenamefont {Chen},\ and\ \citenamefont {Fan}}]{dong2024molecular}%
  \BibitemOpen
  \bibfield  {author} {\bibinfo {author} {\bibfnamefont {H.}~\bibnamefont {Dong}}, \bibinfo {author} {\bibfnamefont {Y.}~\bibnamefont {Shi}}, \bibinfo {author} {\bibfnamefont {P.}~\bibnamefont {Ying}}, \bibinfo {author} {\bibfnamefont {K.}~\bibnamefont {Xu}}, \bibinfo {author} {\bibfnamefont {T.}~\bibnamefont {Liang}}, \bibinfo {author} {\bibfnamefont {Y.}~\bibnamefont {Wang}}, \bibinfo {author} {\bibfnamefont {Z.}~\bibnamefont {Zeng}}, \bibinfo {author} {\bibfnamefont {X.}~\bibnamefont {Wu}}, \bibinfo {author} {\bibfnamefont {W.}~\bibnamefont {Zhou}}, \bibinfo {author} {\bibfnamefont {S.}~\bibnamefont {Xiong}}, \bibinfo {author} {\bibfnamefont {S.}~\bibnamefont {Chen}},\ and\ \bibinfo {author} {\bibfnamefont {Z.}~\bibnamefont {Fan}},\ }\bibfield  {title} {\bibinfo {title} {{Molecular dynamics simulations of heat transport using machine-learned potentials: A mini-review and tutorial on GPUMD with neuroevolution potentials}},\ }\href {https://doi.org/10.1063/5.0200833} {\bibfield  {journal} {\bibinfo
  {journal} {Journal of Applied Physics}\ }\textbf {\bibinfo {volume} {135}},\ \bibinfo {pages} {161101} (\bibinfo {year} {2024})}\BibitemShut {NoStop}%
\bibitem [{\citenamefont {Ying}\ \emph {et~al.}(2025{\natexlab{a}})\citenamefont {Ying}, \citenamefont {Qian}, \citenamefont {Zhao}, \citenamefont {Wang}, \citenamefont {Xu}, \citenamefont {Ding}, \citenamefont {Chen},\ and\ \citenamefont {Fan}}]{ying2025advances}%
  \BibitemOpen
  \bibfield  {author} {\bibinfo {author} {\bibfnamefont {P.}~\bibnamefont {Ying}}, \bibinfo {author} {\bibfnamefont {C.}~\bibnamefont {Qian}}, \bibinfo {author} {\bibfnamefont {R.}~\bibnamefont {Zhao}}, \bibinfo {author} {\bibfnamefont {Y.}~\bibnamefont {Wang}}, \bibinfo {author} {\bibfnamefont {K.}~\bibnamefont {Xu}}, \bibinfo {author} {\bibfnamefont {F.}~\bibnamefont {Ding}}, \bibinfo {author} {\bibfnamefont {S.}~\bibnamefont {Chen}},\ and\ \bibinfo {author} {\bibfnamefont {Z.}~\bibnamefont {Fan}},\ }\bibfield  {title} {\bibinfo {title} {Advances in modeling complex materials: The rise of neuroevolution potentials},\ }\href {https://doi.org/10.1063/5.0259061} {\bibfield  {journal} {\bibinfo  {journal} {Chemical Physics Reviews}\ }\textbf {\bibinfo {volume} {6}},\ \bibinfo {pages} {011310} (\bibinfo {year} {2025}{\natexlab{a}})}\BibitemShut {NoStop}%
\bibitem [{\citenamefont {Xu}\ \emph {et~al.}(2025)\citenamefont {Xu}, \citenamefont {Bu}, \citenamefont {Pan}, \citenamefont {Lindgren}, \citenamefont {Wu}, \citenamefont {Wang}, \citenamefont {Liu}, \citenamefont {Song}, \citenamefont {Xu}, \citenamefont {Li}, \citenamefont {Hainer}, \citenamefont {Svensson}, \citenamefont {Wiktor}, \citenamefont {Zhao}, \citenamefont {Huang}, \citenamefont {Qian}, \citenamefont {Zhang}, \citenamefont {Zeng}, \citenamefont {Zhang}, \citenamefont {Tang}, \citenamefont {Xiao}, \citenamefont {Yan}, \citenamefont {Shi}, \citenamefont {Liang}, \citenamefont {Wang}, \citenamefont {Liang}, \citenamefont {Cao}, \citenamefont {Wang}, \citenamefont {Ying}, \citenamefont {Xu}, \citenamefont {Chen}, \citenamefont {Zhang}, \citenamefont {Chen}, \citenamefont {Wu}, \citenamefont {Jiang}, \citenamefont {Berger}, \citenamefont {Li}, \citenamefont {Chen}, \citenamefont {Gabourie}, \citenamefont {Dong}, \citenamefont {Xiong}, \citenamefont {Wei}, \citenamefont {Chen}, \citenamefont {Xu},
  \citenamefont {Ding}, \citenamefont {Sun}, \citenamefont {Ala-Nissila}, \citenamefont {Harju}, \citenamefont {Zheng}, \citenamefont {Guan}, \citenamefont {Erhart}, \citenamefont {Sun}, \citenamefont {Ouyang}, \citenamefont {Su},\ and\ \citenamefont {Fan}}]{xu2025mega}%
  \BibitemOpen
  \bibfield  {author} {\bibinfo {author} {\bibfnamefont {K.}~\bibnamefont {Xu}}, \bibinfo {author} {\bibfnamefont {H.}~\bibnamefont {Bu}}, \bibinfo {author} {\bibfnamefont {S.}~\bibnamefont {Pan}}, \bibinfo {author} {\bibfnamefont {E.}~\bibnamefont {Lindgren}}, \bibinfo {author} {\bibfnamefont {Y.}~\bibnamefont {Wu}}, \bibinfo {author} {\bibfnamefont {Y.}~\bibnamefont {Wang}}, \bibinfo {author} {\bibfnamefont {J.}~\bibnamefont {Liu}}, \bibinfo {author} {\bibfnamefont {K.}~\bibnamefont {Song}}, \bibinfo {author} {\bibfnamefont {B.}~\bibnamefont {Xu}}, \bibinfo {author} {\bibfnamefont {Y.}~\bibnamefont {Li}}, \bibinfo {author} {\bibfnamefont {T.}~\bibnamefont {Hainer}}, \bibinfo {author} {\bibfnamefont {L.}~\bibnamefont {Svensson}}, \bibinfo {author} {\bibfnamefont {J.}~\bibnamefont {Wiktor}}, \bibinfo {author} {\bibfnamefont {R.}~\bibnamefont {Zhao}}, \bibinfo {author} {\bibfnamefont {H.}~\bibnamefont {Huang}}, \bibinfo {author} {\bibfnamefont {C.}~\bibnamefont {Qian}}, \bibinfo {author} {\bibfnamefont
  {S.}~\bibnamefont {Zhang}}, \bibinfo {author} {\bibfnamefont {Z.}~\bibnamefont {Zeng}}, \bibinfo {author} {\bibfnamefont {B.}~\bibnamefont {Zhang}}, \bibinfo {author} {\bibfnamefont {B.}~\bibnamefont {Tang}}, \bibinfo {author} {\bibfnamefont {Y.}~\bibnamefont {Xiao}}, \bibinfo {author} {\bibfnamefont {Z.}~\bibnamefont {Yan}}, \bibinfo {author} {\bibfnamefont {J.}~\bibnamefont {Shi}}, \bibinfo {author} {\bibfnamefont {Z.}~\bibnamefont {Liang}}, \bibinfo {author} {\bibfnamefont {J.}~\bibnamefont {Wang}}, \bibinfo {author} {\bibfnamefont {T.}~\bibnamefont {Liang}}, \bibinfo {author} {\bibfnamefont {S.}~\bibnamefont {Cao}}, \bibinfo {author} {\bibfnamefont {Y.}~\bibnamefont {Wang}}, \bibinfo {author} {\bibfnamefont {P.}~\bibnamefont {Ying}}, \bibinfo {author} {\bibfnamefont {N.}~\bibnamefont {Xu}}, \bibinfo {author} {\bibfnamefont {C.}~\bibnamefont {Chen}}, \bibinfo {author} {\bibfnamefont {Y.}~\bibnamefont {Zhang}}, \bibinfo {author} {\bibfnamefont {Z.}~\bibnamefont {Chen}}, \bibinfo {author} {\bibfnamefont
  {X.}~\bibnamefont {Wu}}, \bibinfo {author} {\bibfnamefont {W.}~\bibnamefont {Jiang}}, \bibinfo {author} {\bibfnamefont {E.}~\bibnamefont {Berger}}, \bibinfo {author} {\bibfnamefont {Y.}~\bibnamefont {Li}}, \bibinfo {author} {\bibfnamefont {S.}~\bibnamefont {Chen}}, \bibinfo {author} {\bibfnamefont {A.~J.}\ \bibnamefont {Gabourie}}, \bibinfo {author} {\bibfnamefont {H.}~\bibnamefont {Dong}}, \bibinfo {author} {\bibfnamefont {S.}~\bibnamefont {Xiong}}, \bibinfo {author} {\bibfnamefont {N.}~\bibnamefont {Wei}}, \bibinfo {author} {\bibfnamefont {Y.}~\bibnamefont {Chen}}, \bibinfo {author} {\bibfnamefont {J.}~\bibnamefont {Xu}}, \bibinfo {author} {\bibfnamefont {F.}~\bibnamefont {Ding}}, \bibinfo {author} {\bibfnamefont {Z.}~\bibnamefont {Sun}}, \bibinfo {author} {\bibfnamefont {T.}~\bibnamefont {Ala-Nissila}}, \bibinfo {author} {\bibfnamefont {A.}~\bibnamefont {Harju}}, \bibinfo {author} {\bibfnamefont {J.}~\bibnamefont {Zheng}}, \bibinfo {author} {\bibfnamefont {P.}~\bibnamefont {Guan}}, \bibinfo {author}
  {\bibfnamefont {P.}~\bibnamefont {Erhart}}, \bibinfo {author} {\bibfnamefont {J.}~\bibnamefont {Sun}}, \bibinfo {author} {\bibfnamefont {W.}~\bibnamefont {Ouyang}}, \bibinfo {author} {\bibfnamefont {Y.}~\bibnamefont {Su}},\ and\ \bibinfo {author} {\bibfnamefont {Z.}~\bibnamefont {Fan}},\ }\bibfield  {title} {\bibinfo {title} {{GPUMD 4.0: A high-performance molecular dynamics package for versatile materials simulations with machine-learned potentials}},\ }\href {https://doi.org/doi: 10.1002/mgea.70028} {\bibfield  {journal} {\bibinfo  {journal} {Materials Genome Engineering Advances}\ }\textbf {\bibinfo {volume} {3}},\ \bibinfo {pages} {e70028} (\bibinfo {year} {2025})}\BibitemShut {NoStop}%
\bibitem [{\citenamefont {Hockney}\ and\ \citenamefont {Eastwood}(1988)}]{hockney1988computer}%
  \BibitemOpen
  \bibfield  {author} {\bibinfo {author} {\bibfnamefont {R.}~\bibnamefont {Hockney}}\ and\ \bibinfo {author} {\bibfnamefont {J.}~\bibnamefont {Eastwood}},\ }\href {https://doi.org/10.1201/9780367806934} {\emph {\bibinfo {title} {Computer simulation using particles}}}\ (\bibinfo  {publisher} {Taylor \& Francis, Inc.},\ \bibinfo {year} {1988})\BibitemShut {NoStop}%
\bibitem [{\citenamefont {Fan}\ \emph {et~al.}(2021)\citenamefont {Fan}, \citenamefont {Zeng}, \citenamefont {Zhang}, \citenamefont {Wang}, \citenamefont {Song}, \citenamefont {Dong}, \citenamefont {Chen},\ and\ \citenamefont {Ala-Nissila}}]{fan2021neuroevolution}%
  \BibitemOpen
  \bibfield  {author} {\bibinfo {author} {\bibfnamefont {Z.}~\bibnamefont {Fan}}, \bibinfo {author} {\bibfnamefont {Z.}~\bibnamefont {Zeng}}, \bibinfo {author} {\bibfnamefont {C.}~\bibnamefont {Zhang}}, \bibinfo {author} {\bibfnamefont {Y.}~\bibnamefont {Wang}}, \bibinfo {author} {\bibfnamefont {K.}~\bibnamefont {Song}}, \bibinfo {author} {\bibfnamefont {H.}~\bibnamefont {Dong}}, \bibinfo {author} {\bibfnamefont {Y.}~\bibnamefont {Chen}},\ and\ \bibinfo {author} {\bibfnamefont {T.}~\bibnamefont {Ala-Nissila}},\ }\bibfield  {title} {\bibinfo {title} {Neuroevolution machine learning potentials: Combining high accuracy and low cost in atomistic simulations and application to heat transport},\ }\href {https://doi.org/10.1103/PhysRevB.104.104309} {\bibfield  {journal} {\bibinfo  {journal} {Physical Review B}\ }\textbf {\bibinfo {volume} {104}},\ \bibinfo {pages} {104309} (\bibinfo {year} {2021})}\BibitemShut {NoStop}%
\bibitem [{\citenamefont {Fan}(2022)}]{fan2022improving}%
  \BibitemOpen
  \bibfield  {author} {\bibinfo {author} {\bibfnamefont {Z.}~\bibnamefont {Fan}},\ }\bibfield  {title} {\bibinfo {title} {Improving the accuracy of the neuroevolution machine learning potential for multi-component systems},\ }\href {https://doi.org/10.1088/1361-648X/ac462b} {\bibfield  {journal} {\bibinfo  {journal} {Journal of Physics: Condensed Matter}\ }\textbf {\bibinfo {volume} {34}},\ \bibinfo {pages} {125902} (\bibinfo {year} {2022})}\BibitemShut {NoStop}%
\bibitem [{\citenamefont {Fan}\ \emph {et~al.}(2022)\citenamefont {Fan}, \citenamefont {Wang}, \citenamefont {Ying}, \citenamefont {Song}, \citenamefont {Wang}, \citenamefont {Wang}, \citenamefont {Zeng}, \citenamefont {Xu}, \citenamefont {Lindgren}, \citenamefont {Rahm}, \citenamefont {Gabourie}, \citenamefont {Liu}, \citenamefont {Dong}, \citenamefont {Wu}, \citenamefont {Chen}, \citenamefont {Zhong}, \citenamefont {Sun}, \citenamefont {Erhart}, \citenamefont {Su},\ and\ \citenamefont {Ala-Nissila}}]{fan2022gpumd}%
  \BibitemOpen
  \bibfield  {author} {\bibinfo {author} {\bibfnamefont {Z.}~\bibnamefont {Fan}}, \bibinfo {author} {\bibfnamefont {Y.}~\bibnamefont {Wang}}, \bibinfo {author} {\bibfnamefont {P.}~\bibnamefont {Ying}}, \bibinfo {author} {\bibfnamefont {K.}~\bibnamefont {Song}}, \bibinfo {author} {\bibfnamefont {J.}~\bibnamefont {Wang}}, \bibinfo {author} {\bibfnamefont {Y.}~\bibnamefont {Wang}}, \bibinfo {author} {\bibfnamefont {Z.}~\bibnamefont {Zeng}}, \bibinfo {author} {\bibfnamefont {K.}~\bibnamefont {Xu}}, \bibinfo {author} {\bibfnamefont {E.}~\bibnamefont {Lindgren}}, \bibinfo {author} {\bibfnamefont {J.~M.}\ \bibnamefont {Rahm}}, \bibinfo {author} {\bibfnamefont {A.~J.}\ \bibnamefont {Gabourie}}, \bibinfo {author} {\bibfnamefont {J.}~\bibnamefont {Liu}}, \bibinfo {author} {\bibfnamefont {H.}~\bibnamefont {Dong}}, \bibinfo {author} {\bibfnamefont {J.}~\bibnamefont {Wu}}, \bibinfo {author} {\bibfnamefont {Y.}~\bibnamefont {Chen}}, \bibinfo {author} {\bibfnamefont {Z.}~\bibnamefont {Zhong}}, \bibinfo {author}
  {\bibfnamefont {J.}~\bibnamefont {Sun}}, \bibinfo {author} {\bibfnamefont {P.}~\bibnamefont {Erhart}}, \bibinfo {author} {\bibfnamefont {Y.}~\bibnamefont {Su}},\ and\ \bibinfo {author} {\bibfnamefont {T.}~\bibnamefont {Ala-Nissila}},\ }\bibfield  {title} {\bibinfo {title} {{GPUMD: A package for constructing accurate machine-learned potentials and performing highly efficient atomistic simulations}},\ }\href {https://doi.org/10.1063/5.0106617} {\bibfield  {journal} {\bibinfo  {journal} {The Journal of Chemical Physics}\ }\textbf {\bibinfo {volume} {157}},\ \bibinfo {pages} {114801} (\bibinfo {year} {2022})}\BibitemShut {NoStop}%
\bibitem [{\citenamefont {Song}\ \emph {et~al.}(2024{\natexlab{b}})\citenamefont {Song}, \citenamefont {Zhao}, \citenamefont {Liu}, \citenamefont {Wang}, \citenamefont {Lindgren}, \citenamefont {Wang}, \citenamefont {Chen}, \citenamefont {Xu}, \citenamefont {Liang}, \citenamefont {Ying}, \citenamefont {Xu}, \citenamefont {Zhao}, \citenamefont {Shi}, \citenamefont {Wang}, \citenamefont {Lyu}, \citenamefont {Zeng}, \citenamefont {Liang}, \citenamefont {Dong}, \citenamefont {Sun}, \citenamefont {Chen}, \citenamefont {Zhang}, \citenamefont {Guo}, \citenamefont {Qian}, \citenamefont {Sun}, \citenamefont {Erhart}, \citenamefont {Ala-Nissila}, \citenamefont {Su},\ and\ \citenamefont {Fan}}]{song2024general}%
  \BibitemOpen
  \bibfield  {author} {\bibinfo {author} {\bibfnamefont {K.}~\bibnamefont {Song}}, \bibinfo {author} {\bibfnamefont {R.}~\bibnamefont {Zhao}}, \bibinfo {author} {\bibfnamefont {J.}~\bibnamefont {Liu}}, \bibinfo {author} {\bibfnamefont {Y.}~\bibnamefont {Wang}}, \bibinfo {author} {\bibfnamefont {E.}~\bibnamefont {Lindgren}}, \bibinfo {author} {\bibfnamefont {Y.}~\bibnamefont {Wang}}, \bibinfo {author} {\bibfnamefont {S.}~\bibnamefont {Chen}}, \bibinfo {author} {\bibfnamefont {K.}~\bibnamefont {Xu}}, \bibinfo {author} {\bibfnamefont {T.}~\bibnamefont {Liang}}, \bibinfo {author} {\bibfnamefont {P.}~\bibnamefont {Ying}}, \bibinfo {author} {\bibfnamefont {N.}~\bibnamefont {Xu}}, \bibinfo {author} {\bibfnamefont {Z.}~\bibnamefont {Zhao}}, \bibinfo {author} {\bibfnamefont {J.}~\bibnamefont {Shi}}, \bibinfo {author} {\bibfnamefont {J.}~\bibnamefont {Wang}}, \bibinfo {author} {\bibfnamefont {S.}~\bibnamefont {Lyu}}, \bibinfo {author} {\bibfnamefont {Z.}~\bibnamefont {Zeng}}, \bibinfo {author} {\bibfnamefont
  {S.}~\bibnamefont {Liang}}, \bibinfo {author} {\bibfnamefont {H.}~\bibnamefont {Dong}}, \bibinfo {author} {\bibfnamefont {L.}~\bibnamefont {Sun}}, \bibinfo {author} {\bibfnamefont {Y.}~\bibnamefont {Chen}}, \bibinfo {author} {\bibfnamefont {Z.}~\bibnamefont {Zhang}}, \bibinfo {author} {\bibfnamefont {W.}~\bibnamefont {Guo}}, \bibinfo {author} {\bibfnamefont {P.}~\bibnamefont {Qian}}, \bibinfo {author} {\bibfnamefont {J.}~\bibnamefont {Sun}}, \bibinfo {author} {\bibfnamefont {P.}~\bibnamefont {Erhart}}, \bibinfo {author} {\bibfnamefont {T.}~\bibnamefont {Ala-Nissila}}, \bibinfo {author} {\bibfnamefont {Y.}~\bibnamefont {Su}},\ and\ \bibinfo {author} {\bibfnamefont {Z.}~\bibnamefont {Fan}},\ }\bibfield  {title} {\bibinfo {title} {General-purpose machine-learned potential for 16 elemental metals and their alloys},\ }\href {https://doi.org/10.1038/s41467-024-54554-x} {\bibfield  {journal} {\bibinfo  {journal} {Nature Communications}\ }\textbf {\bibinfo {volume} {15}},\ \bibinfo {pages} {10208} (\bibinfo {year}
  {2024}{\natexlab{b}})}\BibitemShut {NoStop}%
\bibitem [{\citenamefont {Schaul}\ \emph {et~al.}(2011)\citenamefont {Schaul}, \citenamefont {Glasmachers},\ and\ \citenamefont {Schmidhuber}}]{schaul2011high}%
  \BibitemOpen
  \bibfield  {author} {\bibinfo {author} {\bibfnamefont {T.}~\bibnamefont {Schaul}}, \bibinfo {author} {\bibfnamefont {T.}~\bibnamefont {Glasmachers}},\ and\ \bibinfo {author} {\bibfnamefont {J.}~\bibnamefont {Schmidhuber}},\ }\bibfield  {title} {\bibinfo {title} {High dimensions and heavy tails for natural evolution strategies},\ }in\ \href {https://doi.org/10.1145/2001576.2001692} {\emph {\bibinfo {booktitle} {Proceedings of the 13th Annual Conference on Genetic and Evolutionary Computation}}},\ \bibinfo {series and number} {GECCO '11}\ (\bibinfo  {publisher} {Association for Computing Machinery},\ \bibinfo {address} {New York, NY, USA},\ \bibinfo {year} {2011})\ p.\ \bibinfo {pages} {845–852}\BibitemShut {NoStop}%
\bibitem [{\citenamefont {Behler}(2011)}]{behler2011atom}%
  \BibitemOpen
  \bibfield  {author} {\bibinfo {author} {\bibfnamefont {J.}~\bibnamefont {Behler}},\ }\bibfield  {title} {\bibinfo {title} {Atom-centered symmetry functions for constructing high-dimensional neural network potentials},\ }\href {https://doi.org/10.1063/1.3553717} {\bibfield  {journal} {\bibinfo  {journal} {The Journal of Chemical Physics}\ }\textbf {\bibinfo {volume} {134}},\ \bibinfo {pages} {074106} (\bibinfo {year} {2011})}\BibitemShut {NoStop}%
\bibitem [{\citenamefont {Lindgren}\ \emph {et~al.}(2024)\citenamefont {Lindgren}, \citenamefont {Rahm}, \citenamefont {Fransson}, \citenamefont {Eriksson}, \citenamefont {{\"O}sterbacka}, \citenamefont {Fan},\ and\ \citenamefont {Erhart}}]{LinRahFra24}%
  \BibitemOpen
  \bibfield  {author} {\bibinfo {author} {\bibfnamefont {E.}~\bibnamefont {Lindgren}}, \bibinfo {author} {\bibfnamefont {M.}~\bibnamefont {Rahm}}, \bibinfo {author} {\bibfnamefont {E.}~\bibnamefont {Fransson}}, \bibinfo {author} {\bibfnamefont {F.}~\bibnamefont {Eriksson}}, \bibinfo {author} {\bibfnamefont {N.}~\bibnamefont {{\"O}sterbacka}}, \bibinfo {author} {\bibfnamefont {Z.}~\bibnamefont {Fan}},\ and\ \bibinfo {author} {\bibfnamefont {P.}~\bibnamefont {Erhart}},\ }\bibfield  {title} {\bibinfo {title} {Calorine: {A Python} package for constructing and sampling neuroevolution potential models},\ }\href {https://doi.org/10.21105/joss.06264} {\bibfield  {journal} {\bibinfo  {journal} {Journal of Open Source Software}\ }\textbf {\bibinfo {volume} {9}},\ \bibinfo {pages} {6264} (\bibinfo {year} {2024})}\BibitemShut {NoStop}%
\bibitem [{\citenamefont {Toukmaji}\ and\ \citenamefont {Board}(1996)}]{toukmaji1996ewald}%
  \BibitemOpen
  \bibfield  {author} {\bibinfo {author} {\bibfnamefont {A.~Y.}\ \bibnamefont {Toukmaji}}\ and\ \bibinfo {author} {\bibfnamefont {J.~A.}\ \bibnamefont {Board}},\ }\bibfield  {title} {\bibinfo {title} {Ewald summation techniques in perspective: a survey},\ }\href {https://doi.org/10.1016/0010-4655(96)00016-1} {\bibfield  {journal} {\bibinfo  {journal} {Computer Physics Communications}\ }\textbf {\bibinfo {volume} {95}},\ \bibinfo {pages} {73} (\bibinfo {year} {1996})}\BibitemShut {NoStop}%
\bibitem [{\citenamefont {Fan}\ \emph {et~al.}(2015)\citenamefont {Fan}, \citenamefont {Pereira}, \citenamefont {Wang}, \citenamefont {Zheng}, \citenamefont {Donadio},\ and\ \citenamefont {Harju}}]{fan2015force}%
  \BibitemOpen
  \bibfield  {author} {\bibinfo {author} {\bibfnamefont {Z.}~\bibnamefont {Fan}}, \bibinfo {author} {\bibfnamefont {L.~F.~C.}\ \bibnamefont {Pereira}}, \bibinfo {author} {\bibfnamefont {H.-Q.}\ \bibnamefont {Wang}}, \bibinfo {author} {\bibfnamefont {J.-C.}\ \bibnamefont {Zheng}}, \bibinfo {author} {\bibfnamefont {D.}~\bibnamefont {Donadio}},\ and\ \bibinfo {author} {\bibfnamefont {A.}~\bibnamefont {Harju}},\ }\bibfield  {title} {\bibinfo {title} {Force and heat current formulas for many-body potentials in molecular dynamics simulations with applications to thermal conductivity calculations},\ }\href {https://doi.org/10.1103/PhysRevB.92.094301} {\bibfield  {journal} {\bibinfo  {journal} {Physical Review B}\ }\textbf {\bibinfo {volume} {92}},\ \bibinfo {pages} {094301} (\bibinfo {year} {2015})}\BibitemShut {NoStop}%
\bibitem [{\citenamefont {Heyes}(1994)}]{Heyes1994prb}%
  \BibitemOpen
  \bibfield  {author} {\bibinfo {author} {\bibfnamefont {D.~M.}\ \bibnamefont {Heyes}},\ }\bibfield  {title} {\bibinfo {title} {Pressure tensor of partial-charge and point-dipole lattices with bulk and surface geometries},\ }\href {https://doi.org/10.1103/PhysRevB.49.755} {\bibfield  {journal} {\bibinfo  {journal} {Physical Review B}\ }\textbf {\bibinfo {volume} {49}},\ \bibinfo {pages} {755} (\bibinfo {year} {1994})}\BibitemShut {NoStop}%
\bibitem [{\citenamefont {Kirby}\ and\ \citenamefont {Jungwirth}(2019)}]{Kirby2019jpcl}%
  \BibitemOpen
  \bibfield  {author} {\bibinfo {author} {\bibfnamefont {B.~J.}\ \bibnamefont {Kirby}}\ and\ \bibinfo {author} {\bibfnamefont {P.}~\bibnamefont {Jungwirth}},\ }\bibfield  {title} {\bibinfo {title} {Charge scaling manifesto: A way of reconciling the inherently macroscopic and microscopic natures of molecular simulations},\ }\href {https://doi.org/10.1021/acs.jpclett.9b02652} {\bibfield  {journal} {\bibinfo  {journal} {The Journal of Physical Chemistry Letters}\ }\textbf {\bibinfo {volume} {10}},\ \bibinfo {pages} {7531} (\bibinfo {year} {2019})}\BibitemShut {NoStop}%
\bibitem [{\citenamefont {Allen}\ and\ \citenamefont {Tildesley}(2017)}]{Allen2017book}%
  \BibitemOpen
  \bibfield  {author} {\bibinfo {author} {\bibfnamefont {M.~P.}\ \bibnamefont {Allen}}\ and\ \bibinfo {author} {\bibfnamefont {D.~J.}\ \bibnamefont {Tildesley}},\ }\href {https://doi.org/10.1093/oso/9780198803195.001.0001} {\emph {\bibinfo {title} {Computer Simulation of Liquids}}}\ (\bibinfo  {publisher} {Oxford University Press},\ \bibinfo {year} {2017})\BibitemShut {NoStop}%
\bibitem [{\citenamefont {Darden}\ \emph {et~al.}(1993)\citenamefont {Darden}, \citenamefont {York},\ and\ \citenamefont {Pedersen}}]{Darden1993jcp}%
  \BibitemOpen
  \bibfield  {author} {\bibinfo {author} {\bibfnamefont {T.}~\bibnamefont {Darden}}, \bibinfo {author} {\bibfnamefont {D.}~\bibnamefont {York}},\ and\ \bibinfo {author} {\bibfnamefont {L.}~\bibnamefont {Pedersen}},\ }\bibfield  {title} {\bibinfo {title} {{Particle mesh Ewald: An $N \log(N)$ method for Ewald sums in large systems}},\ }\href {https://doi.org/10.1063/1.464397} {\bibfield  {journal} {\bibinfo  {journal} {The Journal of Chemical Physics}\ }\textbf {\bibinfo {volume} {98}},\ \bibinfo {pages} {10089} (\bibinfo {year} {1993})}\BibitemShut {NoStop}%
\bibitem [{\citenamefont {Essmann}\ \emph {et~al.}(1995)\citenamefont {Essmann}, \citenamefont {Perera}, \citenamefont {Berkowitz}, \citenamefont {Darden}, \citenamefont {Lee},\ and\ \citenamefont {Pedersen}}]{Essmann1995jcp}%
  \BibitemOpen
  \bibfield  {author} {\bibinfo {author} {\bibfnamefont {U.}~\bibnamefont {Essmann}}, \bibinfo {author} {\bibfnamefont {L.}~\bibnamefont {Perera}}, \bibinfo {author} {\bibfnamefont {M.~L.}\ \bibnamefont {Berkowitz}}, \bibinfo {author} {\bibfnamefont {T.}~\bibnamefont {Darden}}, \bibinfo {author} {\bibfnamefont {H.}~\bibnamefont {Lee}},\ and\ \bibinfo {author} {\bibfnamefont {L.~G.}\ \bibnamefont {Pedersen}},\ }\bibfield  {title} {\bibinfo {title} {{A smooth particle mesh Ewald method}},\ }\href {https://doi.org/10.1063/1.470117} {\bibfield  {journal} {\bibinfo  {journal} {The Journal of Chemical Physics}\ }\textbf {\bibinfo {volume} {103}},\ \bibinfo {pages} {8577} (\bibinfo {year} {1995})}\BibitemShut {NoStop}%
\bibitem [{\citenamefont {Deserno}\ and\ \citenamefont {Holm}(1998)}]{Deserno1998jcp}%
  \BibitemOpen
  \bibfield  {author} {\bibinfo {author} {\bibfnamefont {M.}~\bibnamefont {Deserno}}\ and\ \bibinfo {author} {\bibfnamefont {C.}~\bibnamefont {Holm}},\ }\bibfield  {title} {\bibinfo {title} {{How to mesh up Ewald sums. I. A theoretical and numerical comparison of various particle mesh routines}},\ }\href {https://doi.org/10.1063/1.477414} {\bibfield  {journal} {\bibinfo  {journal} {The Journal of Chemical Physics}\ }\textbf {\bibinfo {volume} {109}},\ \bibinfo {pages} {7678} (\bibinfo {year} {1998})}\BibitemShut {NoStop}%
\bibitem [{\citenamefont {Ballenegger}\ \emph {et~al.}(2012)\citenamefont {Ballenegger}, \citenamefont {Cerdà},\ and\ \citenamefont {Holm}}]{Ballenegger2012jctc}%
  \BibitemOpen
  \bibfield  {author} {\bibinfo {author} {\bibfnamefont {V.}~\bibnamefont {Ballenegger}}, \bibinfo {author} {\bibfnamefont {J.~J.}\ \bibnamefont {Cerdà}},\ and\ \bibinfo {author} {\bibfnamefont {C.}~\bibnamefont {Holm}},\ }\bibfield  {title} {\bibinfo {title} {How to convert {SPME} to {P3M}: {Influence} functions and error estimates},\ }\href {https://doi.org/10.1021/ct2001792} {\bibfield  {journal} {\bibinfo  {journal} {Journal of Chemical Theory and Computation}\ }\textbf {\bibinfo {volume} {8}},\ \bibinfo {pages} {936} (\bibinfo {year} {2012})}\BibitemShut {NoStop}%
\bibitem [{\citenamefont {Xu}\ \emph {et~al.}(2023)\citenamefont {Xu}, \citenamefont {Hao}, \citenamefont {Liang}, \citenamefont {Ying}, \citenamefont {Xu}, \citenamefont {Wu},\ and\ \citenamefont {Fan}}]{XuHaoLia23}%
  \BibitemOpen
  \bibfield  {author} {\bibinfo {author} {\bibfnamefont {K.}~\bibnamefont {Xu}}, \bibinfo {author} {\bibfnamefont {Y.}~\bibnamefont {Hao}}, \bibinfo {author} {\bibfnamefont {T.}~\bibnamefont {Liang}}, \bibinfo {author} {\bibfnamefont {P.}~\bibnamefont {Ying}}, \bibinfo {author} {\bibfnamefont {J.}~\bibnamefont {Xu}}, \bibinfo {author} {\bibfnamefont {J.}~\bibnamefont {Wu}},\ and\ \bibinfo {author} {\bibfnamefont {Z.}~\bibnamefont {Fan}},\ }\bibfield  {title} {\bibinfo {title} {Accurate prediction of heat conductivity of water by a neuroevolution potential},\ }\href {https://doi.org/10.1063/5.0147039} {\bibfield  {journal} {\bibinfo  {journal} {The Journal of Chemical Physics}\ }\textbf {\bibinfo {volume} {158}},\ \bibinfo {pages} {204114} (\bibinfo {year} {2023})}\BibitemShut {NoStop}%
\bibitem [{\citenamefont {Xu}\ \emph {et~al.}(2024)\citenamefont {Xu}, \citenamefont {Rosander}, \citenamefont {Schäfer}, \citenamefont {Lindgren}, \citenamefont {Österbacka}, \citenamefont {Fang}, \citenamefont {Chen}, \citenamefont {He}, \citenamefont {Fan},\ and\ \citenamefont {Erhart}}]{XuRosSch24}%
  \BibitemOpen
  \bibfield  {author} {\bibinfo {author} {\bibfnamefont {N.}~\bibnamefont {Xu}}, \bibinfo {author} {\bibfnamefont {P.}~\bibnamefont {Rosander}}, \bibinfo {author} {\bibfnamefont {C.}~\bibnamefont {Schäfer}}, \bibinfo {author} {\bibfnamefont {E.}~\bibnamefont {Lindgren}}, \bibinfo {author} {\bibfnamefont {N.}~\bibnamefont {Österbacka}}, \bibinfo {author} {\bibfnamefont {M.}~\bibnamefont {Fang}}, \bibinfo {author} {\bibfnamefont {W.}~\bibnamefont {Chen}}, \bibinfo {author} {\bibfnamefont {Y.}~\bibnamefont {He}}, \bibinfo {author} {\bibfnamefont {Z.}~\bibnamefont {Fan}},\ and\ \bibinfo {author} {\bibfnamefont {P.}~\bibnamefont {Erhart}},\ }\bibfield  {title} {\bibinfo {title} {Tensorial properties via the neuroevolution potential framework: Fast simulation of infrared and raman spectra},\ }\href {https://doi.org/10.1021/acs.jctc.3c01343} {\bibfield  {journal} {\bibinfo  {journal} {Journal of Chemical Theory and Computation}\ }\textbf {\bibinfo {volume} {20}},\ \bibinfo {pages} {3273} (\bibinfo {year}
  {2024})}\BibitemShut {NoStop}%
\bibitem [{\citenamefont {Zhang}\ \emph {et~al.}(2021)\citenamefont {Zhang}, \citenamefont {Wang}, \citenamefont {Car},\ and\ \citenamefont {E}}]{ZhaWanCar21}%
  \BibitemOpen
  \bibfield  {author} {\bibinfo {author} {\bibfnamefont {L.}~\bibnamefont {Zhang}}, \bibinfo {author} {\bibfnamefont {H.}~\bibnamefont {Wang}}, \bibinfo {author} {\bibfnamefont {R.}~\bibnamefont {Car}},\ and\ \bibinfo {author} {\bibfnamefont {W.}~\bibnamefont {E}},\ }\bibfield  {title} {\bibinfo {title} {Phase diagram of a deep potential water model},\ }\href {https://doi.org/10.1103/PhysRevLett.126.236001} {\bibfield  {journal} {\bibinfo  {journal} {Physical Review Letters}\ }\textbf {\bibinfo {volume} {126}},\ \bibinfo {pages} {236001} (\bibinfo {year} {2021})}\BibitemShut {NoStop}%
\bibitem [{\citenamefont {Sun}\ \emph {et~al.}(2015)\citenamefont {Sun}, \citenamefont {Ruzsinszky},\ and\ \citenamefont {Perdew}}]{SunRuzPer15}%
  \BibitemOpen
  \bibfield  {author} {\bibinfo {author} {\bibfnamefont {J.}~\bibnamefont {Sun}}, \bibinfo {author} {\bibfnamefont {A.}~\bibnamefont {Ruzsinszky}},\ and\ \bibinfo {author} {\bibfnamefont {J.~P.}\ \bibnamefont {Perdew}},\ }\bibfield  {title} {\bibinfo {title} {Strongly constrained and appropriately normed semilocal density functional},\ }\href {https://doi.org/10.1103/PhysRevLett.115.036402} {\bibfield  {journal} {\bibinfo  {journal} {Physical Review Letters}\ }\textbf {\bibinfo {volume} {115}},\ \bibinfo {pages} {036402} (\bibinfo {year} {2015})}\BibitemShut {NoStop}%
\bibitem [{\citenamefont {Kresse}\ and\ \citenamefont {Furthmüller}(1996)}]{KreFur96a}%
  \BibitemOpen
  \bibfield  {author} {\bibinfo {author} {\bibfnamefont {G.}~\bibnamefont {Kresse}}\ and\ \bibinfo {author} {\bibfnamefont {J.}~\bibnamefont {Furthmüller}},\ }\bibfield  {title} {\bibinfo {title} {Efficiency of ab-initio total energy calculations for metals and semiconductors using a plane-wave basis set},\ }\href {https://doi.org/10.1016/0927-0256(96)00008-0} {\bibfield  {journal} {\bibinfo  {journal} {Computational Materials Science}\ }\textbf {\bibinfo {volume} {6}},\ \bibinfo {pages} {15} (\bibinfo {year} {1996})}\BibitemShut {NoStop}%
\bibitem [{\citenamefont {Bl\"{o}chl}(1994)}]{Blo94}%
  \BibitemOpen
  \bibfield  {author} {\bibinfo {author} {\bibfnamefont {P.~E.}\ \bibnamefont {Bl\"{o}chl}},\ }\bibfield  {title} {\bibinfo {title} {Projector augmented-wave method},\ }\href {https://doi.org/10.1103/PhysRevB.50.17953} {\bibfield  {journal} {\bibinfo  {journal} {Physical Review B}\ }\textbf {\bibinfo {volume} {50}},\ \bibinfo {pages} {17953} (\bibinfo {year} {1994})}\BibitemShut {NoStop}%
\bibitem [{\citenamefont {Kresse}\ and\ \citenamefont {Joubert}(1999)}]{KreJou99}%
  \BibitemOpen
  \bibfield  {author} {\bibinfo {author} {\bibfnamefont {G.}~\bibnamefont {Kresse}}\ and\ \bibinfo {author} {\bibfnamefont {D.}~\bibnamefont {Joubert}},\ }\bibfield  {title} {\bibinfo {title} {From ultrasoft pseudopotentials to the projector augmented-wave method},\ }\href {https://doi.org/10.1103/PhysRevB.59.1758} {\bibfield  {journal} {\bibinfo  {journal} {Physical Review B}\ }\textbf {\bibinfo {volume} {59}},\ \bibinfo {pages} {1758} (\bibinfo {year} {1999})}\BibitemShut {NoStop}%
\bibitem [{\citenamefont {Thormählen}\ \emph {et~al.}(1985)\citenamefont {Thormählen}, \citenamefont {Straub},\ and\ \citenamefont {Grigull}}]{refractive_index_water_1985}%
  \BibitemOpen
  \bibfield  {author} {\bibinfo {author} {\bibfnamefont {I.}~\bibnamefont {Thormählen}}, \bibinfo {author} {\bibfnamefont {J.}~\bibnamefont {Straub}},\ and\ \bibinfo {author} {\bibfnamefont {U.}~\bibnamefont {Grigull}},\ }\bibfield  {title} {\bibinfo {title} {Refractive index of water and its dependence on wavelength, temperature, and density},\ }\href {https://doi.org/10.1063/1.555743} {\bibfield  {journal} {\bibinfo  {journal} {Journal of Physical and Chemical Reference Data}\ }\textbf {\bibinfo {volume} {14}},\ \bibinfo {pages} {933} (\bibinfo {year} {1985})}\BibitemShut {NoStop}%
\bibitem [{\citenamefont {Harvey}\ \emph {et~al.}(1998)\citenamefont {Harvey}, \citenamefont {Gallagher},\ and\ \citenamefont {Sengers}}]{refractive_index_water_1998}%
  \BibitemOpen
  \bibfield  {author} {\bibinfo {author} {\bibfnamefont {A.~H.}\ \bibnamefont {Harvey}}, \bibinfo {author} {\bibfnamefont {J.~S.}\ \bibnamefont {Gallagher}},\ and\ \bibinfo {author} {\bibfnamefont {J.~M. H.~L.}\ \bibnamefont {Sengers}},\ }\bibfield  {title} {\bibinfo {title} {Revised formulation for the refractive index of water and steam as a function of wavelength, temperature and density},\ }\href {https://doi.org/10.1063/1.556029} {\bibfield  {journal} {\bibinfo  {journal} {Journal of Physical and Chemical Reference Data}\ }\textbf {\bibinfo {volume} {27}},\ \bibinfo {pages} {761} (\bibinfo {year} {1998})}\BibitemShut {NoStop}%
\bibitem [{\citenamefont {Lemmon}\ \emph {et~al.}(2026)\citenamefont {Lemmon}, \citenamefont {Bell}, \citenamefont {Huber},\ and\ \citenamefont {McLinden}}]{NIST_water}%
  \BibitemOpen
  \bibfield  {author} {\bibinfo {author} {\bibfnamefont {E.~W.}\ \bibnamefont {Lemmon}}, \bibinfo {author} {\bibfnamefont {I.~H.}\ \bibnamefont {Bell}}, \bibinfo {author} {\bibfnamefont {M.~L.}\ \bibnamefont {Huber}},\ and\ \bibinfo {author} {\bibfnamefont {M.~O.}\ \bibnamefont {McLinden}},\ }\bibfield  {title} {\bibinfo {title} {Thermophysical properties of fluid systems},\ }in\ \href {https://doi.org/10.18434/T4D303} {\emph {\bibinfo {booktitle} {NIST Chemistry WebBook, NIST Standard Reference Database Number 69}}},\ \bibinfo {editor} {edited by\ \bibinfo {editor} {\bibfnamefont {P.~J.}\ \bibnamefont {Linstrom}}\ and\ \bibinfo {editor} {\bibfnamefont {W.~G.}\ \bibnamefont {Mallard}}}\ (\bibinfo  {publisher} {National Institute of Standards and Technology},\ \bibinfo {address} {Gaithersburg MD, 20899},\ \bibinfo {year} {retrieved January 11, 2026})\BibitemShut {NoStop}%
\bibitem [{\citenamefont {Ying}\ \emph {et~al.}(2025{\natexlab{b}})\citenamefont {Ying}, \citenamefont {Zhou}, \citenamefont {Svensson}, \citenamefont {Berger}, \citenamefont {Fransson}, \citenamefont {Eriksson}, \citenamefont {Xu}, \citenamefont {Liang}, \citenamefont {Xu}, \citenamefont {Song}, \citenamefont {Chen}, \citenamefont {Erhart},\ and\ \citenamefont {Fan}}]{YinZhoSve25}%
  \BibitemOpen
  \bibfield  {author} {\bibinfo {author} {\bibfnamefont {P.}~\bibnamefont {Ying}}, \bibinfo {author} {\bibfnamefont {W.}~\bibnamefont {Zhou}}, \bibinfo {author} {\bibfnamefont {L.}~\bibnamefont {Svensson}}, \bibinfo {author} {\bibfnamefont {E.}~\bibnamefont {Berger}}, \bibinfo {author} {\bibfnamefont {E.}~\bibnamefont {Fransson}}, \bibinfo {author} {\bibfnamefont {F.}~\bibnamefont {Eriksson}}, \bibinfo {author} {\bibfnamefont {K.}~\bibnamefont {Xu}}, \bibinfo {author} {\bibfnamefont {T.}~\bibnamefont {Liang}}, \bibinfo {author} {\bibfnamefont {J.}~\bibnamefont {Xu}}, \bibinfo {author} {\bibfnamefont {B.}~\bibnamefont {Song}}, \bibinfo {author} {\bibfnamefont {S.}~\bibnamefont {Chen}}, \bibinfo {author} {\bibfnamefont {P.}~\bibnamefont {Erhart}},\ and\ \bibinfo {author} {\bibfnamefont {Z.}~\bibnamefont {Fan}},\ }\bibfield  {title} {\bibinfo {title} {Highly efficient path-integral molecular dynamics simulations with gpumd using neuroevolution potentials: Case studies on thermal properties of materials},\ }\href
  {https://doi.org/10.1063/5.0241006} {\bibfield  {journal} {\bibinfo  {journal} {Journal of Chemical Physics}\ }\textbf {\bibinfo {volume} {162}},\ \bibinfo {pages} {064109} (\bibinfo {year} {2025}{\natexlab{b}})}\BibitemShut {NoStop}%
\bibitem [{\citenamefont {Downing}\ and\ \citenamefont {Williams}(1975)}]{DowWil75}%
  \BibitemOpen
  \bibfield  {author} {\bibinfo {author} {\bibfnamefont {H.~D.}\ \bibnamefont {Downing}}\ and\ \bibinfo {author} {\bibfnamefont {D.}~\bibnamefont {Williams}},\ }\bibfield  {title} {\bibinfo {title} {Optical constants of water in the infrared},\ }\href {https://doi.org/10.1029/JC080i012p01656} {\bibfield  {journal} {\bibinfo  {journal} {Journal of Geophysical Research}\ }\textbf {\bibinfo {volume} {80}},\ \bibinfo {pages} {1656} (\bibinfo {year} {1975})}\BibitemShut {NoStop}%
\bibitem [{\citenamefont {Max}\ and\ \citenamefont {Chapados}(2009)}]{MaxCha09}%
  \BibitemOpen
  \bibfield  {author} {\bibinfo {author} {\bibfnamefont {J.-J.}\ \bibnamefont {Max}}\ and\ \bibinfo {author} {\bibfnamefont {C.}~\bibnamefont {Chapados}},\ }\bibfield  {title} {\bibinfo {title} {Isotope effects in liquid water by infrared spectroscopy. {III}. \ce{H2O} and \ce{D2O} spectra from 6000 to 0 cm$^{-1}$},\ }\href {https://doi.org/10.1063/1.3258646} {\bibfield  {journal} {\bibinfo  {journal} {Journal of Chemical Physics}\ }\textbf {\bibinfo {volume} {131}},\ \bibinfo {pages} {184505} (\bibinfo {year} {2009})}\BibitemShut {NoStop}%
\bibitem [{\citenamefont {Chen}\ \emph {et~al.}(2015)\citenamefont {Chen}, \citenamefont {Rangasamy}, \citenamefont {dela Cruz}, \citenamefont {Liang},\ and\ \citenamefont {An}}]{chen2015study}%
  \BibitemOpen
  \bibfield  {author} {\bibinfo {author} {\bibfnamefont {Y.}~\bibnamefont {Chen}}, \bibinfo {author} {\bibfnamefont {E.}~\bibnamefont {Rangasamy}}, \bibinfo {author} {\bibfnamefont {C.~R.}\ \bibnamefont {dela Cruz}}, \bibinfo {author} {\bibfnamefont {C.}~\bibnamefont {Liang}},\ and\ \bibinfo {author} {\bibfnamefont {K.}~\bibnamefont {An}},\ }\bibfield  {title} {\bibinfo {title} {A study of suppressed formation of low-conductivity phases in doped \ce{Li7La3Zr2O_{12}} garnets by in situ neutron diffraction},\ }\href {https://doi.org/10.1039/C5TA04902D} {\bibfield  {journal} {\bibinfo  {journal} {Journal of Materials Chemistry A}\ }\textbf {\bibinfo {volume} {3}},\ \bibinfo {pages} {22868} (\bibinfo {year} {2015})}\BibitemShut {NoStop}%
\bibitem [{\citenamefont {Han}\ \emph {et~al.}(2016)\citenamefont {Han}, \citenamefont {Zhu}, \citenamefont {He}, \citenamefont {Mo},\ and\ \citenamefont {Wang}}]{han2016electrochemical}%
  \BibitemOpen
  \bibfield  {author} {\bibinfo {author} {\bibfnamefont {F.}~\bibnamefont {Han}}, \bibinfo {author} {\bibfnamefont {Y.}~\bibnamefont {Zhu}}, \bibinfo {author} {\bibfnamefont {X.}~\bibnamefont {He}}, \bibinfo {author} {\bibfnamefont {Y.}~\bibnamefont {Mo}},\ and\ \bibinfo {author} {\bibfnamefont {C.}~\bibnamefont {Wang}},\ }\bibfield  {title} {\bibinfo {title} {Electrochemical stability of \ce{Li_{10}GeP_2S_{12}} and \ce{Li7La3Zr2O_{12}} solid electrolytes},\ }\href {https://doi.org/10.1002/aenm.201501590} {\bibfield  {journal} {\bibinfo  {journal} {Advanced Energy Materials}\ }\textbf {\bibinfo {volume} {6}},\ \bibinfo {pages} {1501590} (\bibinfo {year} {2016})}\BibitemShut {NoStop}%
\bibitem [{\citenamefont {Yan}\ and\ \citenamefont {Zhu}(2024)}]{yan2024impact}%
  \BibitemOpen
  \bibfield  {author} {\bibinfo {author} {\bibfnamefont {Z.}~\bibnamefont {Yan}}\ and\ \bibinfo {author} {\bibfnamefont {Y.}~\bibnamefont {Zhu}},\ }\bibfield  {title} {\bibinfo {title} {Impact of lithium nonstoichiometry on ionic diffusion in tetragonal garnet-type \ce{Li7La3Zr2O_{12}}},\ }\href {https://doi.org/10.1021/acs.chemmater.4c02454} {\bibfield  {journal} {\bibinfo  {journal} {Chemistry of Materials}\ }\textbf {\bibinfo {volume} {36}},\ \bibinfo {pages} {11551} (\bibinfo {year} {2024})}\BibitemShut {NoStop}%
\bibitem [{\citenamefont {Perdew}\ \emph {et~al.}(2008)\citenamefont {Perdew}, \citenamefont {Ruzsinszky}, \citenamefont {Csonka}, \citenamefont {Vydrov}, \citenamefont {Scuseria}, \citenamefont {Constantin}, \citenamefont {Zhou},\ and\ \citenamefont {Burke}}]{Perdew2008_PBEsol}%
  \BibitemOpen
  \bibfield  {author} {\bibinfo {author} {\bibfnamefont {J.~P.}\ \bibnamefont {Perdew}}, \bibinfo {author} {\bibfnamefont {A.}~\bibnamefont {Ruzsinszky}}, \bibinfo {author} {\bibfnamefont {G.~I.}\ \bibnamefont {Csonka}}, \bibinfo {author} {\bibfnamefont {O.~A.}\ \bibnamefont {Vydrov}}, \bibinfo {author} {\bibfnamefont {G.~E.}\ \bibnamefont {Scuseria}}, \bibinfo {author} {\bibfnamefont {L.~A.}\ \bibnamefont {Constantin}}, \bibinfo {author} {\bibfnamefont {X.}~\bibnamefont {Zhou}},\ and\ \bibinfo {author} {\bibfnamefont {K.}~\bibnamefont {Burke}},\ }\bibfield  {title} {\bibinfo {title} {Restoring the density-gradient expansion for exchange in solids and surfaces},\ }\href {https://doi.org/10.1103/PhysRevLett.100.136406} {\bibfield  {journal} {\bibinfo  {journal} {Physical Review Letters}\ }\textbf {\bibinfo {volume} {100}},\ \bibinfo {pages} {136406} (\bibinfo {year} {2008})}\BibitemShut {NoStop}%
\bibitem [{\citenamefont {Martyna}\ \emph {et~al.}(1996)\citenamefont {Martyna}, \citenamefont {Tuckerman}, \citenamefont {Tobias},\ and\ \citenamefont {Klein}}]{MarTucTob96}%
  \BibitemOpen
  \bibfield  {author} {\bibinfo {author} {\bibfnamefont {G.~J.}\ \bibnamefont {Martyna}}, \bibinfo {author} {\bibfnamefont {M.~E.}\ \bibnamefont {Tuckerman}}, \bibinfo {author} {\bibfnamefont {D.~J.}\ \bibnamefont {Tobias}},\ and\ \bibinfo {author} {\bibfnamefont {M.~L.}\ \bibnamefont {Klein}},\ }\bibfield  {title} {\bibinfo {title} {Explicit reversible integrators for extended systems dynamics},\ }\href {https://doi.org/10.1080/00268979600100761} {\bibfield  {journal} {\bibinfo  {journal} {Molecular Physics}\ }\textbf {\bibinfo {volume} {87}},\ \bibinfo {pages} {1117} (\bibinfo {year} {1996})}\BibitemShut {NoStop}%
\bibitem [{\citenamefont {Merz}(1949)}]{Merz1949}%
  \BibitemOpen
  \bibfield  {author} {\bibinfo {author} {\bibfnamefont {W.~J.}\ \bibnamefont {Merz}},\ }\bibfield  {title} {\bibinfo {title} {The electric and optical behavior of \ce{BaTiO3} single-domain crystals},\ }\href {https://doi.org/10.1103/PhysRev.76.1221} {\bibfield  {journal} {\bibinfo  {journal} {Physical Review}\ }\textbf {\bibinfo {volume} {76}},\ \bibinfo {pages} {1221} (\bibinfo {year} {1949})}\BibitemShut {NoStop}%
\bibitem [{\citenamefont {Benedict}\ and\ \citenamefont {Durand}(1958)}]{Benedict1958}%
  \BibitemOpen
  \bibfield  {author} {\bibinfo {author} {\bibfnamefont {T.~S.}\ \bibnamefont {Benedict}}\ and\ \bibinfo {author} {\bibfnamefont {J.~L.}\ \bibnamefont {Durand}},\ }\bibfield  {title} {\bibinfo {title} {Dielectric properties of single domain crystals of \ce{BaTiO3} at microwave frequencies},\ }\href {https://doi.org/10.1103/PhysRev.109.1091} {\bibfield  {journal} {\bibinfo  {journal} {Physical Review}\ }\textbf {\bibinfo {volume} {109}},\ \bibinfo {pages} {1091} (\bibinfo {year} {1958})}\BibitemShut {NoStop}%
\bibitem [{\citenamefont {Wieder}(1955)}]{Wieder1955}%
  \BibitemOpen
  \bibfield  {author} {\bibinfo {author} {\bibfnamefont {H.~H.}\ \bibnamefont {Wieder}},\ }\bibfield  {title} {\bibinfo {title} {Electrical behavior of barium titanate single crystals at low temperatures},\ }\href {https://doi.org/10.1103/PhysRev.99.1161} {\bibfield  {journal} {\bibinfo  {journal} {Physical Review}\ }\textbf {\bibinfo {volume} {99}},\ \bibinfo {pages} {1161} (\bibinfo {year} {1955})}\BibitemShut {NoStop}%
\bibitem [{\citenamefont {Wang}\ and\ \citenamefont {Lai}(2015)}]{WanLai15}%
  \BibitemOpen
  \bibfield  {author} {\bibinfo {author} {\bibfnamefont {Y.}~\bibnamefont {Wang}}\ and\ \bibinfo {author} {\bibfnamefont {W.}~\bibnamefont {Lai}},\ }\bibfield  {title} {\bibinfo {title} {Phase transition in lithium garnet oxide ionic conductors \ce{Li7La3Zr2O_{12}}: {{The}} role of {{Ta}} substitution and \ce{H2O}/\ce{CO2} exposure},\ }\href {https://doi.org/10.1016/j.jpowsour.2014.11.062} {\bibfield  {journal} {\bibinfo  {journal} {Journal of Power Sources}\ }\textbf {\bibinfo {volume} {275}},\ \bibinfo {pages} {612} (\bibinfo {year} {2015})}\BibitemShut {NoStop}%
\bibitem [{\citenamefont {Fransson}\ \emph {et~al.}(2021)\citenamefont {Fransson}, \citenamefont {Slabanja}, \citenamefont {Erhart},\ and\ \citenamefont {Wahnström}}]{dynasor1}%
  \BibitemOpen
  \bibfield  {author} {\bibinfo {author} {\bibfnamefont {E.}~\bibnamefont {Fransson}}, \bibinfo {author} {\bibfnamefont {M.}~\bibnamefont {Slabanja}}, \bibinfo {author} {\bibfnamefont {P.}~\bibnamefont {Erhart}},\ and\ \bibinfo {author} {\bibfnamefont {G.}~\bibnamefont {Wahnström}},\ }\bibfield  {title} {\bibinfo {title} {dynasor--a tool for extracting dynamical structure factors and current correlation functions from molecular dynamics simulations},\ }\href {https://doi.org/10.1002/adts.202000240} {\bibfield  {journal} {\bibinfo  {journal} {Advanced Theory and Simulations}\ }\textbf {\bibinfo {volume} {4}},\ \bibinfo {pages} {2000240} (\bibinfo {year} {2021})}\BibitemShut {NoStop}%
\bibitem [{\citenamefont {Berger}\ \emph {et~al.}(2025)\citenamefont {Berger}, \citenamefont {Fransson}, \citenamefont {Eriksson}, \citenamefont {Lindgren}, \citenamefont {Wahnström}, \citenamefont {Rod},\ and\ \citenamefont {Erhart}}]{dynasor2}%
  \BibitemOpen
  \bibfield  {author} {\bibinfo {author} {\bibfnamefont {E.}~\bibnamefont {Berger}}, \bibinfo {author} {\bibfnamefont {E.}~\bibnamefont {Fransson}}, \bibinfo {author} {\bibfnamefont {F.}~\bibnamefont {Eriksson}}, \bibinfo {author} {\bibfnamefont {E.}~\bibnamefont {Lindgren}}, \bibinfo {author} {\bibfnamefont {G.}~\bibnamefont {Wahnström}}, \bibinfo {author} {\bibfnamefont {T.~H.}\ \bibnamefont {Rod}},\ and\ \bibinfo {author} {\bibfnamefont {P.}~\bibnamefont {Erhart}},\ }\bibfield  {title} {\bibinfo {title} {Dynasor 2: From simulation to experiment through correlation functions},\ }\href {https://doi.org/10.1016/j.cpc.2025.109759} {\bibfield  {journal} {\bibinfo  {journal} {Computer Physics Communications}\ }\textbf {\bibinfo {volume} {316}},\ \bibinfo {pages} {109759} (\bibinfo {year} {2025})}\BibitemShut {NoStop}%
\bibitem [{\citenamefont {Gigli}\ \emph {et~al.}(2022)\citenamefont {Gigli}, \citenamefont {Veit}, \citenamefont {Kotiuga}, \citenamefont {Pizzi}, \citenamefont {Marzari},\ and\ \citenamefont {Ceriotti}}]{Gigli2022}%
  \BibitemOpen
  \bibfield  {author} {\bibinfo {author} {\bibfnamefont {L.}~\bibnamefont {Gigli}}, \bibinfo {author} {\bibfnamefont {M.}~\bibnamefont {Veit}}, \bibinfo {author} {\bibfnamefont {M.}~\bibnamefont {Kotiuga}}, \bibinfo {author} {\bibfnamefont {G.}~\bibnamefont {Pizzi}}, \bibinfo {author} {\bibfnamefont {N.}~\bibnamefont {Marzari}},\ and\ \bibinfo {author} {\bibfnamefont {M.}~\bibnamefont {Ceriotti}},\ }\bibfield  {title} {\bibinfo {title} {Thermodynamics and dielectric response of \ce{BaTiO3} by data-driven modeling},\ }\href {https://doi.org/10.1038/s41524-022-00845-0} {\bibfield  {journal} {\bibinfo  {journal} {npj Computational Materials}\ }\textbf {\bibinfo {volume} {8}},\ \bibinfo {pages} {209} (\bibinfo {year} {2022})}\BibitemShut {NoStop}%
\bibitem [{\citenamefont {Lindgren}\ \emph {et~al.}(2025)\citenamefont {Lindgren}, \citenamefont {Jackson}, \citenamefont {Fransson}, \citenamefont {Berger}, \citenamefont {Rudi\'c}, \citenamefont {\v{S}koro}, \citenamefont {Turanyi}, \citenamefont {Mukhopadhyay},\ and\ \citenamefont {Erhart}}]{LinJacFra25}%
  \BibitemOpen
  \bibfield  {author} {\bibinfo {author} {\bibfnamefont {E.}~\bibnamefont {Lindgren}}, \bibinfo {author} {\bibfnamefont {A.~J.}\ \bibnamefont {Jackson}}, \bibinfo {author} {\bibfnamefont {E.}~\bibnamefont {Fransson}}, \bibinfo {author} {\bibfnamefont {E.}~\bibnamefont {Berger}}, \bibinfo {author} {\bibfnamefont {S.}~\bibnamefont {Rudi\'c}}, \bibinfo {author} {\bibfnamefont {G.}~\bibnamefont {\v{S}koro}}, \bibinfo {author} {\bibfnamefont {R.}~\bibnamefont {Turanyi}}, \bibinfo {author} {\bibfnamefont {S.}~\bibnamefont {Mukhopadhyay}},\ and\ \bibinfo {author} {\bibfnamefont {P.}~\bibnamefont {Erhart}},\ }\bibfield  {title} {\bibinfo {title} {Predicting neutron experiments from first principles: A workflow powered by machine learning},\ }\href {https://doi.org/10.1039/D5TA03325J} {\bibfield  {journal} {\bibinfo  {journal} {Journal of Materials Chemistry A}\ }\textbf {\bibinfo {volume} {13}},\ \bibinfo {pages} {25509} (\bibinfo {year} {2025})}\BibitemShut {NoStop}%
\bibitem [{\citenamefont {Furness}\ \emph {et~al.}(2020)\citenamefont {Furness}, \citenamefont {Kaplan}, \citenamefont {Ning}, \citenamefont {Perdew},\ and\ \citenamefont {Sun}}]{Furness2020}%
  \BibitemOpen
  \bibfield  {author} {\bibinfo {author} {\bibfnamefont {J.~W.}\ \bibnamefont {Furness}}, \bibinfo {author} {\bibfnamefont {A.~D.}\ \bibnamefont {Kaplan}}, \bibinfo {author} {\bibfnamefont {J.}~\bibnamefont {Ning}}, \bibinfo {author} {\bibfnamefont {J.~P.}\ \bibnamefont {Perdew}},\ and\ \bibinfo {author} {\bibfnamefont {J.}~\bibnamefont {Sun}},\ }\bibfield  {title} {\bibinfo {title} {Accurate and numerically efficient {r2SCAN} meta-generalized gradient approximation},\ }\href {https://doi.org/10.1021/acs.jpclett.0c02405} {\bibfield  {journal} {\bibinfo  {journal} {The Journal of Physical Chemistry Letters}\ }\textbf {\bibinfo {volume} {11}},\ \bibinfo {pages} {8208} (\bibinfo {year} {2020})}\BibitemShut {NoStop}%
\bibitem [{\citenamefont {Hashimoto}\ and\ \citenamefont {Moriwake}(2015)}]{Hashimoto2015}%
  \BibitemOpen
  \bibfield  {author} {\bibinfo {author} {\bibfnamefont {T.}~\bibnamefont {Hashimoto}}\ and\ \bibinfo {author} {\bibfnamefont {H.}~\bibnamefont {Moriwake}},\ }\bibfield  {title} {\bibinfo {title} {Dielectric properties of \ce{BaTiO3} by molecular dynamics simulations using a shell model},\ }\href {https://doi.org/10.1080/08927022.2014.938067} {\bibfield  {journal} {\bibinfo  {journal} {Molecular Simulation}\ }\textbf {\bibinfo {volume} {41}},\ \bibinfo {pages} {1074} (\bibinfo {year} {2015})}\BibitemShut {NoStop}%
\bibitem [{\citenamefont {Jiang}\ \emph {et~al.}(2022)\citenamefont {Jiang}, \citenamefont {Parsonnet}, \citenamefont {Qualls}, \citenamefont {Zhao}, \citenamefont {Susarla}, \citenamefont {Pesquera}, \citenamefont {Dasgupta}, \citenamefont {Acharya}, \citenamefont {Zhang}, \citenamefont {Gosavi}, \citenamefont {Lin}, \citenamefont {Nikonov}, \citenamefont {Li}, \citenamefont {Young}, \citenamefont {Ramesh},\ and\ \citenamefont {Martin}}]{Jiang2022}%
  \BibitemOpen
  \bibfield  {author} {\bibinfo {author} {\bibfnamefont {Y.}~\bibnamefont {Jiang}}, \bibinfo {author} {\bibfnamefont {E.}~\bibnamefont {Parsonnet}}, \bibinfo {author} {\bibfnamefont {A.}~\bibnamefont {Qualls}}, \bibinfo {author} {\bibfnamefont {W.}~\bibnamefont {Zhao}}, \bibinfo {author} {\bibfnamefont {S.}~\bibnamefont {Susarla}}, \bibinfo {author} {\bibfnamefont {D.}~\bibnamefont {Pesquera}}, \bibinfo {author} {\bibfnamefont {A.}~\bibnamefont {Dasgupta}}, \bibinfo {author} {\bibfnamefont {M.}~\bibnamefont {Acharya}}, \bibinfo {author} {\bibfnamefont {H.}~\bibnamefont {Zhang}}, \bibinfo {author} {\bibfnamefont {T.}~\bibnamefont {Gosavi}}, \bibinfo {author} {\bibfnamefont {C.-C.}\ \bibnamefont {Lin}}, \bibinfo {author} {\bibfnamefont {D.~E.}\ \bibnamefont {Nikonov}}, \bibinfo {author} {\bibfnamefont {H.}~\bibnamefont {Li}}, \bibinfo {author} {\bibfnamefont {I.~A.}\ \bibnamefont {Young}}, \bibinfo {author} {\bibfnamefont {R.}~\bibnamefont {Ramesh}},\ and\ \bibinfo {author} {\bibfnamefont {L.~W.}\ \bibnamefont
  {Martin}},\ }\bibfield  {title} {\bibinfo {title} {Enabling ultra-low-voltage switching in \ce{BaTiO3}},\ }\href {https://doi.org/10.1038/s41563-022-01266-6} {\bibfield  {journal} {\bibinfo  {journal} {Nature Materials}\ }\textbf {\bibinfo {volume} {21}},\ \bibinfo {pages} {779} (\bibinfo {year} {2022})}\BibitemShut {NoStop}%
\bibitem [{\citenamefont {Shin}\ \emph {et~al.}(2007)\citenamefont {Shin}, \citenamefont {Grinberg}, \citenamefont {Chen},\ and\ \citenamefont {Rappe}}]{ShiGriChe07}%
  \BibitemOpen
  \bibfield  {author} {\bibinfo {author} {\bibfnamefont {Y.-H.}\ \bibnamefont {Shin}}, \bibinfo {author} {\bibfnamefont {I.}~\bibnamefont {Grinberg}}, \bibinfo {author} {\bibfnamefont {I.-W.}\ \bibnamefont {Chen}},\ and\ \bibinfo {author} {\bibfnamefont {A.~M.}\ \bibnamefont {Rappe}},\ }\bibfield  {title} {\bibinfo {title} {Nucleation and growth mechanism of ferroelectric domain-wall motion},\ }\href {https://doi.org/10.1038/nature06165} {\bibfield  {journal} {\bibinfo  {journal} {Nature}\ }\textbf {\bibinfo {volume} {449}},\ \bibinfo {pages} {881} (\bibinfo {year} {2007})}\BibitemShut {NoStop}%
\bibitem [{\citenamefont {Togo}\ \emph {et~al.}(2023)\citenamefont {Togo}, \citenamefont {Chaput}, \citenamefont {Tadano},\ and\ \citenamefont {Tanaka}}]{togo2023implementation}%
  \BibitemOpen
  \bibfield  {author} {\bibinfo {author} {\bibfnamefont {A.}~\bibnamefont {Togo}}, \bibinfo {author} {\bibfnamefont {L.}~\bibnamefont {Chaput}}, \bibinfo {author} {\bibfnamefont {T.}~\bibnamefont {Tadano}},\ and\ \bibinfo {author} {\bibfnamefont {I.}~\bibnamefont {Tanaka}},\ }\bibfield  {title} {\bibinfo {title} {Implementation strategies in phonopy and phono3py},\ }\href {https://doi.org/10.1088/1361-648X/acd831} {\bibfield  {journal} {\bibinfo  {journal} {Journal of Physics: Condensed Matter}\ }\textbf {\bibinfo {volume} {35}},\ \bibinfo {pages} {353001} (\bibinfo {year} {2023})}\BibitemShut {NoStop}%
\bibitem [{\citenamefont {Togo}(2023)}]{phonopy2023}%
  \BibitemOpen
  \bibfield  {author} {\bibinfo {author} {\bibfnamefont {A.}~\bibnamefont {Togo}},\ }\bibfield  {title} {\bibinfo {title} {First-principles phonon calculations with phonopy and phono3py},\ }\href {https://doi.org/10.7566/JPSJ.92.012001} {\bibfield  {journal} {\bibinfo  {journal} {Journal of the Physical Society of Japan}\ }\textbf {\bibinfo {volume} {92}},\ \bibinfo {pages} {012001} (\bibinfo {year} {2023})}\BibitemShut {NoStop}%
\bibitem [{\citenamefont {Thomas}\ \emph {et~al.}(2010)\citenamefont {Thomas}, \citenamefont {Turney}, \citenamefont {Iutzi}, \citenamefont {Amon},\ and\ \citenamefont {McGaughey}}]{Thomas2010}%
  \BibitemOpen
  \bibfield  {author} {\bibinfo {author} {\bibfnamefont {J.~A.}\ \bibnamefont {Thomas}}, \bibinfo {author} {\bibfnamefont {J.~E.}\ \bibnamefont {Turney}}, \bibinfo {author} {\bibfnamefont {R.~M.}\ \bibnamefont {Iutzi}}, \bibinfo {author} {\bibfnamefont {C.~H.}\ \bibnamefont {Amon}},\ and\ \bibinfo {author} {\bibfnamefont {A.~J.~H.}\ \bibnamefont {McGaughey}},\ }\bibfield  {title} {\bibinfo {title} {Predicting phonon dispersion relations and lifetimes from the spectral energy density},\ }\href {https://doi.org/10.1103/PhysRevB.81.081411} {\bibfield  {journal} {\bibinfo  {journal} {Physical Review B}\ }\textbf {\bibinfo {volume} {81}},\ \bibinfo {pages} {081411} (\bibinfo {year} {2010})}\BibitemShut {NoStop}%
\bibitem [{\citenamefont {Benbouzid}\ \emph {et~al.}(2022)\citenamefont {Benbouzid}, \citenamefont {Gomes}, \citenamefont {Costa}, \citenamefont {Gharbi}, \citenamefont {P{\'e}b{\`e}re}, \citenamefont {Rossi}, \citenamefont {Tran}, \citenamefont {Tribollet}, \citenamefont {Turmine},\ and\ \citenamefont {Vivier}}]{BenGomCos22}%
  \BibitemOpen
  \bibfield  {author} {\bibinfo {author} {\bibfnamefont {A.~Z.}\ \bibnamefont {Benbouzid}}, \bibinfo {author} {\bibfnamefont {M.~P.}\ \bibnamefont {Gomes}}, \bibinfo {author} {\bibfnamefont {I.}~\bibnamefont {Costa}}, \bibinfo {author} {\bibfnamefont {O.}~\bibnamefont {Gharbi}}, \bibinfo {author} {\bibfnamefont {N.}~\bibnamefont {P{\'e}b{\`e}re}}, \bibinfo {author} {\bibfnamefont {J.~L.}\ \bibnamefont {Rossi}}, \bibinfo {author} {\bibfnamefont {M.~T.~T.}\ \bibnamefont {Tran}}, \bibinfo {author} {\bibfnamefont {B.}~\bibnamefont {Tribollet}}, \bibinfo {author} {\bibfnamefont {M.}~\bibnamefont {Turmine}},\ and\ \bibinfo {author} {\bibfnamefont {V.}~\bibnamefont {Vivier}},\ }\bibfield  {title} {\bibinfo {title} {A new look on the corrosion mechanism of magnesium: {{An EIS}} investigation at different {{pH}}},\ }\href {https://doi.org/10.1016/j.corsci.2022.110463} {\bibfield  {journal} {\bibinfo  {journal} {Corrosion Science}\ }\textbf {\bibinfo {volume} {205}},\ \bibinfo {pages} {110463} (\bibinfo {year}
  {2022})}\BibitemShut {NoStop}%
\bibitem [{\citenamefont {Liu}\ \emph {et~al.}(2025)\citenamefont {Liu}, \citenamefont {Sha}, \citenamefont {Song}, \citenamefont {Wang},\ and\ \citenamefont {Zhang}}]{liu2025cej}%
  \BibitemOpen
  \bibfield  {author} {\bibinfo {author} {\bibfnamefont {Z.}~\bibnamefont {Liu}}, \bibinfo {author} {\bibfnamefont {J.}~\bibnamefont {Sha}}, \bibinfo {author} {\bibfnamefont {G.-L.}\ \bibnamefont {Song}}, \bibinfo {author} {\bibfnamefont {Z.}~\bibnamefont {Wang}},\ and\ \bibinfo {author} {\bibfnamefont {Y.}~\bibnamefont {Zhang}},\ }\bibfield  {title} {\bibinfo {title} {Understanding magnesium dissolution through machine learning molecular dynamics},\ }\href {https://doi.org/10.1016/j.cej.2025.163578} {\bibfield  {journal} {\bibinfo  {journal} {Chemical Engineering Journal}\ }\textbf {\bibinfo {volume} {516}},\ \bibinfo {pages} {163578} (\bibinfo {year} {2025})}\BibitemShut {NoStop}%
\bibitem [{\citenamefont {Liu}\ \emph {et~al.}(2023)\citenamefont {Liu}, \citenamefont {Bao}, \citenamefont {Sha},\ and\ \citenamefont {Zhang}}]{LiuBaoSha23}%
  \BibitemOpen
  \bibfield  {author} {\bibinfo {author} {\bibfnamefont {Z.}~\bibnamefont {Liu}}, \bibinfo {author} {\bibfnamefont {J.}~\bibnamefont {Bao}}, \bibinfo {author} {\bibfnamefont {J.}~\bibnamefont {Sha}},\ and\ \bibinfo {author} {\bibfnamefont {Z.}~\bibnamefont {Zhang}},\ }\bibfield  {title} {\bibinfo {title} {Modulation of the discharge and corrosion properties of aqueous {Mg}--air batteries by alloying from first-principles theory},\ }\href {https://doi.org/10.1021/acs.jpcc.3c00111} {\bibfield  {journal} {\bibinfo  {journal} {The Journal of Physical Chemistry C}\ }\textbf {\bibinfo {volume} {127}},\ \bibinfo {pages} {10062} (\bibinfo {year} {2023})}\BibitemShut {NoStop}%
\bibitem [{\citenamefont {Yu}\ \emph {et~al.}(2022)\citenamefont {Yu}, \citenamefont {Hong}, \citenamefont {Chen}, \citenamefont {Gong},\ and\ \citenamefont {Xiang}}]{yu2022capturing}%
  \BibitemOpen
  \bibfield  {author} {\bibinfo {author} {\bibfnamefont {H.}~\bibnamefont {Yu}}, \bibinfo {author} {\bibfnamefont {L.}~\bibnamefont {Hong}}, \bibinfo {author} {\bibfnamefont {S.}~\bibnamefont {Chen}}, \bibinfo {author} {\bibfnamefont {X.}~\bibnamefont {Gong}},\ and\ \bibinfo {author} {\bibfnamefont {H.}~\bibnamefont {Xiang}},\ }\href {https://arxiv.org/abs/2211.16684} {\bibinfo {title} {Capturing long-range interaction with reciprocal space neural network}} (\bibinfo {year} {2022}),\ \Eprint {https://arxiv.org/abs/2211.16684} {arXiv:2211.16684 [cond-mat.mtrl-sci]} \BibitemShut {NoStop}%
\bibitem [{\citenamefont {Loche}\ \emph {et~al.}(2025)\citenamefont {Loche}, \citenamefont {Huguenin-Dumittan}, \citenamefont {Honarmand}, \citenamefont {Xu}, \citenamefont {Rumiantsev}, \citenamefont {How}, \citenamefont {Langer},\ and\ \citenamefont {Ceriotti}}]{Loche2025jcp}%
  \BibitemOpen
  \bibfield  {author} {\bibinfo {author} {\bibfnamefont {P.}~\bibnamefont {Loche}}, \bibinfo {author} {\bibfnamefont {K.~K.}\ \bibnamefont {Huguenin-Dumittan}}, \bibinfo {author} {\bibfnamefont {M.}~\bibnamefont {Honarmand}}, \bibinfo {author} {\bibfnamefont {Q.}~\bibnamefont {Xu}}, \bibinfo {author} {\bibfnamefont {E.}~\bibnamefont {Rumiantsev}}, \bibinfo {author} {\bibfnamefont {W.~B.}\ \bibnamefont {How}}, \bibinfo {author} {\bibfnamefont {M.~F.}\ \bibnamefont {Langer}},\ and\ \bibinfo {author} {\bibfnamefont {M.}~\bibnamefont {Ceriotti}},\ }\bibfield  {title} {\bibinfo {title} {Fast and flexible long-range models for atomistic machine learning},\ }\href {https://doi.org/10.1063/5.0251713} {\bibfield  {journal} {\bibinfo  {journal} {The Journal of Chemical Physics}\ }\textbf {\bibinfo {volume} {162}},\ \bibinfo {pages} {142501} (\bibinfo {year} {2025})}\BibitemShut {NoStop}%
\bibitem [{\citenamefont {Ji}\ \emph {et~al.}(2025)\citenamefont {Ji}, \citenamefont {Liang},\ and\ \citenamefont {Xu}}]{ji2025prl}%
  \BibitemOpen
  \bibfield  {author} {\bibinfo {author} {\bibfnamefont {Y.}~\bibnamefont {Ji}}, \bibinfo {author} {\bibfnamefont {J.}~\bibnamefont {Liang}},\ and\ \bibinfo {author} {\bibfnamefont {Z.}~\bibnamefont {Xu}},\ }\bibfield  {title} {\bibinfo {title} {Machine-learning interatomic potentials for long-range systems},\ }\href {https://doi.org/10.1103/ssp9-7s81} {\bibfield  {journal} {\bibinfo  {journal} {Physical Review Letters}\ }\textbf {\bibinfo {volume} {135}},\ \bibinfo {pages} {178001} (\bibinfo {year} {2025})}\BibitemShut {NoStop}%
\end{thebibliography}
\end{document}